\documentclass[review]{elsarticle}
\usepackage{lineno,hyperref,eurosym}
\usepackage{color}
\usepackage{tabularx}
\usepackage[utf8]{inputenc}
\usepackage{graphicx}
\usepackage{amsmath}
\usepackage{lscape}
\usepackage{xcolor, soul}
\usepackage{makecell}

\modulolinenumbers[3]

\journal{PLOS ONE}
\begin{document}
\begin{frontmatter}

\title{Double Gravity-Assist Rendezvous Trajectory to Halley's Comet Using Deep-Space Low Thrust}
\author[label1]{R. Flores}
\author[label1]{A. Beolchi}
\author[label1]{C. Pozzi}
\author[label2]{M. Pontani}
\author[label3]{I. Bertini}
\author[label4,label5]{C. Barbieri}
\author[label1,label6]{E. Fantino$^{\star}$}
\address[label1]{Department of Aerospace Engineering, Khalifa University of Science and Technology, 
P.O. Box 127788, Abu Dhabi (United Arab Emirates)}
\address[label2]{Department of Mechanical and Aerospace Engineering, Sapienza Università di Roma, Rome (Italy)}
\address[label3]{Department of Science and Technology, University of Naples "Parthenope", Naples (Italy)}
\address[label4]{Department of Physics and Astronomy "G. Galilei", University of Padova, Padua (Italy)}
\address[label5]{INAF, Istituto Nazionale di Astrofisica (Italy)}
\address[label6]{Polar Research Center, Khalifa University of Science and Technology, 
P.O. Box 127788, Abu Dhabi (United Arab Emirates)}
\cortext[cor]{Corresponding author: elena.fantino@ku.ac.ae (E.~Fantino)}

\begin{abstract}
The perihelion of comet 1P/Halley in 2061 is an excellent chance to revisit this object of outstanding scientific and cultural relevance. During its 1986 approach to the Sun, it was targeted by several flyby missions. Due to its retrograde, highly-inclined orbit, the relative velocities during the encounter were large, limiting the scientific return due to the short time spent inside the coma. A rendezvous trajectory would overcome this limitation, but the design is challenging due to the limitations of current propulsion technology. Given the lead times of spacecraft development and the long duration of the interplanetary transfer to the comet, it is imperative to start mission planning as soon as possible. We present a novel rendezvous strategy, combining unpowered Jupiter and Saturn gravity-assists with deep-space low-thrust arcs. It minimizes launch energy and propellant budget, constraining the arrival to occur before the onset of high activity. The double flyby strategy reduces the otherwise prohibitive cost of the plane change maneuver. Closed expressions for the optimal flyby geometry, together with an explicit low-thrust transcription technique, reduce the number of design parameter to three, improving computational efficiency.
Crucially, this is the first rendezvous mission concept achievable with well-proven technology (standard radioisotope thermoelectric generators and a Hall-effect thruster) and compatible with several existing launchers. We describe the trajectory optimization strategy and perform a comprehensive exploration of the design space. Finally, we present two promising proof-of-concept trajectories in detail.
\end{abstract}


\end{frontmatter}


\section{Introduction}
\label{sec_introduction}
Comets have long captivated the interest of the general public. Moreover, they are celestial bodies of great scientific importance. It is essential to understand their formation process, how they reached their present reservoirs, what are their constituent materials and the mechanics of their activity. This will help determine the role they played in planetary formation and the appearance of water and life on Earth \citep{Barbieri:2017}.
Comets are categorized into short and long period, with the boundary between the two groups set at a period of 200 years. Compared to the long-period objects, many short-period comets are fainter and less active. Nonetheless, short-period comets are the only ones whose appearance can be predicted accurately, making them ideal targets for exploration missions.

1P/Halley (Halley hereafter) stands out among short-period comets due to its cultural and scientific relevance. It is large and bright, which facilitated naked-eye sightings, observations and records during at least two millennia \citep{Barbieri:2017}. Its orbit is known with high precision, which assisted planning for the flyby missions in 1985/1986 \citep{Brady:1971,Yeomans:1981}. Its next perihelion, in 2061, is expected to attract significant attention from space agencies worldwide, prompting a new wave of close observation attempts. However, Halley presents major challenges for trajectory design due to its retrograde orbit, high ecliptic inclination ($162^\circ$) and eccentricity (0.967), which disfavor prograde intercepts. 

During its latest approach to the Sun in 1986, several space missions encountered Halley at relative velocities of 70-80~$\mathrm{km~s^{-1}}$ \citep{Reinhard:1982,Stelzried:1986,Oya:1986,Sagdeev:1982,Sagdeev:1986,Keller:1986,Reinhard:1987,Hirao:1982}. This resulted in a brief period of close observation within the coma, which limited both the quality and quantity of the collected data \citep{Reinhard:1986,Curdt:1988,Volwerk:2014,Keller:2020}.  
Only one hemisphere of the nucleus could be imaged, and the rotational state could not be accurately measured. Scientists are tempted to explain the nucleus of Halley as a contact binary, but only an extended collection of high-resolution images covering the entire body can confirm this hypothesis. Several other matters require detailed investigation, such as the geomorphology of the surface, the localization and dimension of the active regions, the amount of mass loss, and the internal structure. To address most of these unresolved questions, it is crucial to encounter the comet well before the onset of high activity, at a heliocentric distance larger than the radius of Mars' orbit \citep{Barbieri:2025}.

The above considerations justify the realization of a rendezvous mission, where the spacecraft approaches the comet with very small relative velocity. Designing a trajectory of this kind is complicated, primarily due to the prohibitive cost of reversing the direction of the spacecraft's heliocentric motion and match the orbital energy of the comet. Another hurdle is that the major axis of Halley's orbit is almost perpendicular to its line of nodes. For many short-period comets, these two directions are nearly aligned, making it possible to rendezvous via a three-impulse ballistic trajectory, characterized by a single mid-course maneuver near the line of nodes \citep{Manning:1970}. This plane change impulse can be replaced with a gravity assist (GA) maneuver with a giant planet, in order to save propellant. A rendezvous trajectory to Halley's comet cannot benefit from the three-impulse scheme.

Even though more accessible alternatives existed, like comets 2P/Encke and 22P/Kopff, Halley was considered a priority target for rendezvous missions in the period 1975-1995 \citep{Friedlander:1970,Friedlander:1971,Friedlander:1971b}. These authors explored impulsive trajectories combined with a Jupiter gravity assist maneuver, as well as direct transfers with high-power nuclear electric propulsion ($\sim$100~kW). They concluded that the latter were more promising. At the time, these power plants were expected to become available in the short term. They would have solved the dynamical complications of the mission, enabling large payload mass, low characteristic launch energy, short flight times (a key constraint due to the late start of the research) and wide launch windows. Unfortunately, the technology failed to materialize. Even today, despite it has never been demonstrated in practice,, some viability studies continue to rely on high-power nuclear reactors \citep{Horsewood:2023}.

An alternative to direct rendezvous trajectories with nuclear propulsion is to leverage a GA maneuver with a giant planet, as proposed by Michielsen \cite{Michielsen:1968}. The author combined a Jupiter or Saturn GA with an impulsive trajectory reversal at the aphelion of the post-flyby leg, and a final rendezvous burn. However, the feasibility with the launchers available at the time was questionable due to the high characteristic Earth departure energy \citep{Kruse:1969}. This issue was exacerbated by the late start of the studies, which precluded launch dates with more favorable positions of the giant planets. Mission concepts published in the following years placed more emphasis on nuclear-electric propulsion \citep{Kruse:1969,Friedlander:1970}.

A combination of unpowered Jupiter GA with solar electric propulsion is explored by Horsewood et al. \cite{Horsewood:2023}. However, the concept requires a 40~kW solar array, which is at odds with the proposed dry mass of the spacecraft (1.3 tons). Consider for example the heavier Juno probe \citep{junonasasicence}, which needed 340~kg of solar panels to generate just 12~kW on Earth.

Two recent papers \citep{Beolchi:2024,Barbieri:2025} assumed a ballistic departure from Earth and a single flyby with a giant planet (Jupiter or Saturn), followed by low-thrust arcs to rendezvous with the comet. The concept assumes a dry mas of 1000~kg and 36~mN of thrust, compatible with existing electric propulsion technologies. Unfortunately, the launch energy is extremely high: $C3>150~\mathrm{km^2s^{-2}}$ (where $C3$ denotes the square of the hyperbolic excess speed). The mission would be possible with a super-heavy launcher like the SLS \citep{Stough:2021}, but at an exorbitant cost.

The previous investigations highlight the importance of starting mission planning decades in advance, and evaluating the technological requirements in relation to the state of the art. To reach the comet in time for its 2061 perihelion, the assessment of rendezvous opportunities must commence immediately. This contribution presents a mission concept combining low-thrust (LT) arcs and GA maneuvers with Jupiter and Saturn. The Earth-Jupiter segment uses continuous thrust to increase the mechanical energy of the spacecraft, allowing for a reduction of $C3$. The Jupiter flyby provides an additional boost, further decreasing the launch energy requirement. Next, the Saturn GA inserts the probe into a retrograde trajectory with the correct inclination to intercept the comet. A launch mass between 1500 and 2000 kg is targeted, using a Hall-effect thruster (HET) powered by standard radioisotope thermoelectric generators (RTGs). The departure energy required is compatible with existing heavy launchers.

The trajectory design starts assuming impulsive maneuvers for the Jupiter-Saturn and Saturn-Halley legs. Introducing explicit expressions for the optimal GA geometry, the complete trajectory is parametrized as a function of $C3$ and the dates of Saturn GA and comet rendezvous only. The reduced dimensionality of the problem enables a fast comprehensive exploration of the solution space.

In a second step, the trajectory is refined using LT for all segments, using the impulsive solutions as initial guesses to improve the convergence. An explicit LT transcription method is used to compute the thrust law without altering the dimensionality of the problem. The small number of degrees of freedom, and the high-quality initial approximations from the first step, ensure a fast and robust minimization of the propellant budget. This is, as far as the authors know, the first practical rendezvous trajectory concept that relies only on proven technologies.

This manuscript is organized as follows. Section~\ref{sec_dyn_spac} defines the mathematical model and the assumed performance of the spacecraft. The trajectory design strategy is the subject of Sect.~\ref{sec_tra_des}. The numerical results are presented and discussed in Sect.~\ref{sec_res_dis}. Finally, Sect.~\ref{sec_con} is devoted to the concluding remarks.

\section{Material and methods}
\label{sec_dyn_spac}
\subsection{Dynamical model}
\label{sec_dyn_mod}
At any time, the motion of the spacecraft is affected by the gravity of a single celestial body: the Sun for interplanetary arcs and the relevant planet (Jupiter or Saturn) during the GA maneuvers. The flyby is considered instantaneous, assuming a zero-radius Sphere Of Influence (SOI).

The trajectories of the planets and the comet are computed from high-fidelity ephemeris downloaded from JPL's Solar System Dynamics website \citep{horsys}. All the orbital elements are osculated at the 1 January 2035 epoch and referenced to the heliocentric mean ecliptic and equinox of J2000 reference frame. For reference, Table~\ref{tab_hal} list the elements of the comet (semimajor axis $a$, eccentricity $e$, inclination $i$, longitude of the ascending node $\Omega$, argument of pericenter $\omega$ and epoch of pericenter $t_p$). 

\begin{table}
	\centering
	\caption{Orbital elements of Halley in ecliptic frame. Osculation epoch 1 January 2035.}
	\label{tab_hal}
	\begin{tabular}{cccccc}
		\hline
		$a$ & $e$ & $i$ & $\Omega$ & $\omega$  & $t_p$ \\
		\hline
		17.9 au & 0.967 & $162^\circ$ & $59^\circ.5$ & $112^\circ$ & JD 2474056.1 \\
		\hline
	\end{tabular}
\end{table}

\subsection{Spacecraft characteristics}
\label{sec_spa_cha}
The propulsion system is assumed to be fully steerable, with constant maximum thrust and specific impulse. Because the spacecraft travels toward the outer Solar System and remains far from the Sun for most of the mission duration, RTGs are chosen as the source of electric power. The propulsion system characteristics (maximum thrust $T_{\max}$, specific impulse $I_{sp}$ and input power $P_{in}$) are summarized in Table~\ref{tab:eng}. These performance figures, taken from a design of a tour of the inner moons of Saturn \citep{Fantino:2023}, correspond to the PPS\textcopyright X00 HET \citep{Vaudolon:2019}. While other types of electric motors deliver higher specific impulse, HETs have lower power requirement for a given thrust. This is a critical advantage, in view of the power constraints.

\begin{table}
	\centering
	\caption{Assumed thruster performance parameters.}
	\begin{tabular}{ccc} \hline
		$T_{\max}$ & $I_{sp}$ & $P_{in}$ \\ \hline
		36 mN & 1600 s & 640 W \\ \hline
	\end{tabular}
	\label{tab:eng}
\end{table}

\begin{figure}
	\centering
	\includegraphics[width=0.670\textwidth]{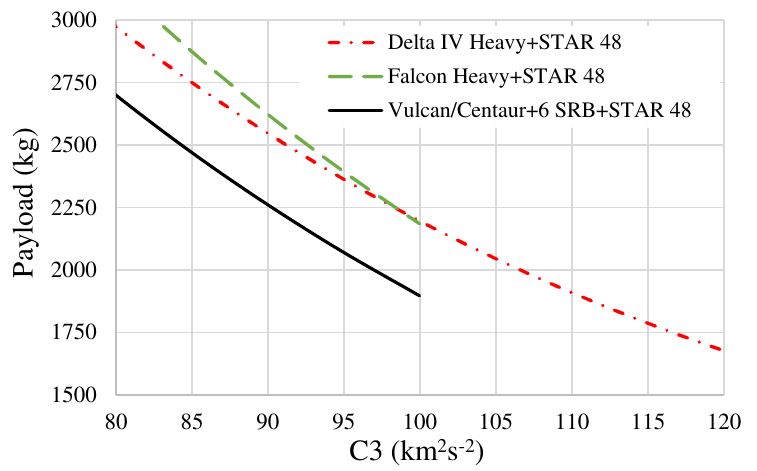}
	\caption{Useful payload vs. departure energy for three heavy-class launchers. Data sourced from \citep{PLASMA}.}
	\label{fig:Launchers}
\end{figure}

\subsection{Launcher performance}
\label{sec_lau_per}
A previous study \citep{Beolchi:2024,Barbieri:2025} assumed a single GA with a giant planet followed by LT arcs to rendezvous with Halley. It required a launch energy above $150~\mathrm{km^2s^{-2}}$ to execute the gravity-assisted plane change. While technically feasible, a super-heavy class launcher would be needed to deliver a reasonable payload to the comet, making the mission cost prohibitive. To overcome this limitation, the launch energy and mass have been constrained to values compatible with existing heavy launchers. Figure~\ref{fig:Launchers} shows the performance of three such platforms, all equipped with a STAR48 kick stage \citep{STAR48}, as a function of $C3$\footnote{No information is available for Falcon Heavy and Vulcan/Centaur for $C3> \mathrm{100~km^2s^{-2}}$, but this does not mean that they are not capable of higher energies.}. The values are taken from the PLASMA website \citep{PLASMA}. In view of the chart, $C3$ was restricted to the interval [80,120]~$\mathrm{km^2s^{-2}}$. The lower bound eliminates departure trajectories that reach their aphelion before intercepting Jupiter's orbit, giving rise to unacceptably long Earth-Jupiter legs. For this $C3$ range, a launch mass of 1500~kg can be assumed without compromising the feasibility of the mission.

\subsection{Trajectory design method}
\label{sec_tra_des}
The objective of this work is to prove the feasibility of the mission, without use of exotic technologies. Therefore, it is not necessary to find an optimal solution, a technically viable trajectory is sufficient to validate the concept. This enables reducing the dimensionality of the problem, simplifying the design process. The trajectory is divided into three stages: 
\begin{itemize}
	\item LT transfer between Earth and Jupiter.
	\item GA with Jupiter and powered transfer to Saturn.
	\item GA with Saturn and powered rendezvous with Halley.
\end{itemize}
The propulsion system operates continuously at maximum thrust during the Earth-Jupiter segment. It increases the spacecraft energy, lowering the $C3$ requirement. To maximize the effectiveness of the propulsion system, the thrust is assumed to act parallel to the spacecraft's heliocentric velocity. This yields the fastest rate of increase of mechanical energy. Furthermore, to maximize the utilization of the launch vehicle's energy, the relative departure velocity from Earth is chosen approximately parallel to the planet's motion. It cannot be exactly parallel, because the orbits of Jupiter and Earth are not coplanar (see Sec.~\ref{sec_seg_12} for more details). Note that, because the magnitude and direction of thrust are prescribed beforehand, the launch conditions determine the complete Earth-Jupiter leg. 

Even using two GAs with the giant planets, the departure energy is substantial ($C3 \sim \mathrm{100~km^2s^{-2}}$). In principle, the launch cost could be lowered via flybys with the inner planets. However, this is impractical because it would consume too much time. Consider, for example, that a resonant flyby with Earth would require at least two years. It will be shown that the time available to prepare the mission is around 10 years for a direct launch to Jupiter. This is already a tight margin, considering the typical mission lead times, so additional flybys have not been considered. For the same reason, past missions requiring large heliocentric energy changes have launched directly with a high $C3$ before performing any GA. Notable examples include Ulysses \citep{Smith1991}, New Horizons \citep{Guo:2008} and Parker Solar Probe \citep{Fox:2016}.

The Jupiter-Saturn and Saturn-Halley legs are computed in two stages. A preliminary step uses impulsive Lambert transfers. All segments are connected with unpowered GA maneuvers. The Saturn and Halley encounter dates are the free parameters, with the GA geometries chosen in a way that minimizes the total impulse. This simple approximation allows for a fast comprehensive exploration of the solution space. Once a promising mission profile is found, the impulsive transfers are replaced with continuous LT arcs, compatible with electric propulsion. The conversion uses an explicit transcription method, which is very efficient computationally. The departure conditions, Saturn flyby date and Halley rendezvous epoch are adjusted to minimize the total propellant budget.

In the following, subscript 1 denotes the departure event, 2 and 3 are the states immediately before/after the Jupiter GA, 4/5 are the pre/post Saturn GA states, and 6 is the rendezvous with the comet. For example, due to the instantaneous GA with zero-radius SOI hypothesis $t_2=t_3$ and ${\bf r}_2={\bf r}_3$ (time and position are continuous) but ${\bf v}_2\neq{\bf v}_3$ (velocity is discontinuous).

Additionally, superscript $\odot$ denotes the Sun, $E$ stands for Earth, $J$ for Jupiter, $S$ for Saturn and $H$ represents Halley. Because the spacecraft must rendezvous with the comet, ${\bf r}_6={\bf r}^H_6$ and ${\bf v}_6={\bf v}^H_6$.

\subsection{Preliminary step: Trajectory design with impulsive maneuvers between Jupiter and Halley}
\label{sec_imp_tra}
All calculations, except the GA maneuvers, use a heliocentric Cartesian ecliptic reference frame $xyz$ with the unit vector $\bf i$ pointing along the direction of the spring equinox, $\bf k$ along the pole of the plane, and $\bf j$ completing the right-handed triad \{${\bf i},{\bf j},{\bf k}$\}.

It will be shown that the launch year ($y_1$), the departure energy ($C3$), and the dates of the Saturn GA ($t_4=t_5$) and Halley redezvous ($t_6$) are sufficient to fully characterize the trajectory. This is possible because the optimal flyby parameters (depth and inclination of the planetocentric hyperbola) can be expressed in terms of these variables. Note that $y_1$ is an integer optimization variable because, for a fixed $C3$, the launch window opens once per Earth-Jupiter synodic period ($\sim$13 months). Thus, for a given departure year, the dates of launch ($t_1$) and Jupiter flyby ($t_2=t_3$) are functions of $C3$ only. This simplifies the optimization problem, because only three design parameters remain: $\{C3, t_5, t_6\}$. 

\subsubsection{Earth-Jupiter segment}
\label{sec_seg_12}
The spacecraft state vector ${\bf x} = [{\bf r}, {\bf v}, m]^T$, where $\bf r$ denotes position, $\bf v$ velocity and $m$ mass, evolves in time according to:
\begin{equation} \label{gov_eq_r}
	\dot{\bf r} = {\bf v},
\end{equation}
\begin{equation} \label{gov_eq_v}
	\dot{\bf v} = -\mu^\odot \frac{\bf r}{r^3} + T_{\max} \frac{\bf v}{v},	
\end{equation}
\begin{equation} \label{gov_eq_m}
	\dot{m} = -\frac{T_{\max}}{I_{sp} g_0},
\end{equation}
for $t \in [t_1, t_2]$. In the equations above, $\mu^\odot$ denotes the gravitational parameter of the Sun and $g_0$ is the standard sea-level acceleration of gravity ($9.81~\mathrm{m~s^{-2}}$). Because the spacecraft departs from Earth and arrives at Jupiter, the conditions ${\bf r}_1={\bf r}(t_1)={\bf r}^E_1$ and ${\bf r}_2={\bf r}^J_2$ must hold, with the positions of the planets computed from the respective ephemeris. Moving forward, to reduce clutter in the notation, the event subscript will be dropped for the celestial bodies, because their state is only relevant when the spacecraft encounters them. That is, ${\bf r}^E={\bf r}^E_1$, ${\bf r}^J={\bf r}^J_2$, and so on.
Because thrust always acts parallel to the velocity vector (Eq.~\ref{gov_eq_v}), the spacecraft trajectory is planar. However, it is not contained in the ecliptic because Jupiter's orbit is inclined. The trajectory plane passes through the Sun, Earth and Jupiter. The normal to this plane is given by:
\begin{equation} \label{eq_n12}
	{\bf n} =  \frac {{\bf r}^E \times {\bf r}^J} {\| {\bf r}^E \times {\bf r}^J \|}.
\end{equation}
Therefore, ${\bf n} \cdot {\bf v}_1 = 0$. Let ${\bf v}^{rel} = {\bf v}_1 - {\bf v}^E$ denote the relative departure velocity, such that $v^{rel} = \sqrt{C3}$. The relative velocity is split into normal (n) and in-plane (b) components
\begin{equation} \label{eq_vnip}
	v^n = -{\bf n} \cdot {\bf v}^E \quad , \quad v^b = \sqrt{C3 - (v^n)^2}.
\end{equation}
The direction $\bf b$ is chosen along the projection of ${\bf v}^E$ over the trajectory plane:
\begin{equation} \label{eq_t}
	{\bf b} = \frac {{\bf v}^E + v^n {\bf n}} {\| {\bf v}^E + v^n {\bf n} \|}.
\end{equation}
The utilization of the launcher's energy is maximized by departing Earth in the direction of $\bf b$
\begin{equation} \label{eq_v1}
	{\bf v}_1 = ({\bf v}^E \cdot {\bf b} + v^b) {\bf b}.
\end{equation}
Because the direction of the initial velocity is prescribed, the trajectory is entirely controlled by the departure energy $C3$ and the launch epoch $t_1$. Moreover, the spacecraft must encounter Jupiter. Thus, for a given $C3$, the Earth-Jupiter phasing is determined. This fixes $t_1$ almost entirely, the only ambiguity being the synodic cycle of the Jupiter-Earth system when the launch takes place. Thus, given the calendar year of launch\footnote{The synodic period of the system is 399 days, so there is one out of twelve calendar years without a launch window.} $y_1$ and $C3$, the problem is uniquely determined. It requires an iterative solution because $t_1$ and $t_2$ are unknown a priory.
The steps to find the solution are as follows:
\begin{enumerate}
	\item Select $C3$ and a guess of $t_1$.
	\item Compute ${\bf v}_1$ with Eq.~\ref{eq_v1} and integrate numerically Eqs.~\ref{gov_eq_r}-\ref{gov_eq_m} from $t_1$ until the spacecraft crosses the orbit of Jupiter ($t_2$).
	\item Compute the difference in ecliptic longitude between the spacecraft and Jupiter at $t_2$. If it is larger than a preset tolerance, correct $t_1$ and return to step 2. 
\end{enumerate}
The solution is found with the secant method, with a tolerance of 1~arcsec for the ecliptic longitude. To ensure a fast and robust convergence, it is desirable to start with a good estimate of the launch date. This is easy to achieve if the approximate time of flight (ToF) $t_2-t_1$ and the transfer angle are known. Because the orbits of the planets are close to circular, these two values depend strongly on $C3$ and weakly on $t_1$. Thus, it is possible to approximate them as functions of $C3$ only and use the mean motion of the planets to estimate when their phasing is adequate. This yields an excellent initial guess of $t_1$. The approximate relations for $ToF$ and transfer angle were obtained fitting a sample of propagated trajectories with generalized least-squares using a B-spline representation based on van Leeuwen \& Fantino \cite{Leeuwen:2003}.

The method described above yields discrete values of $t_1$ for a given $C3$, one for each $y_1$. In real life, it is desirable to have wider launch windows to accommodate deviations from the nominal mission schedule. This can be accomplished varying the orientation of the departure velocity. That is, making the in-plane component of ${\bf v}^{rel}$ not exactly parallel to {\bf b}. As long as the change in direction is small, the total heliocentric energy can be maintained with a minimal impact on $C3$. It is also possible to change the direction of thrust to alter the phasing, giving additional flexibility. However, these considerations are beyond the scope of this work, which focuses on the feasibility of the concept.

\subsubsection{Jupiter-Saturn and Saturn-Halley segments with impulsive maneuvers}
\label{sec_seg_3456}
For each pair of values $\{t_5, t_6\}$, Lambert's problem is solved to connect the points ${\bf r}_5={\bf r}^S$ and ${\bf r}_6 = {\bf r}^H$ with a ToF $t_6-t_5$. Let ${\bf v}^L_5$ and ${\bf v}^L_6$ denote the terminal velocities of the Lambert arc. The two impulses required to rendezvous with the comet using a ballistic trajectory are $\Delta V_5 = \| {\bf v}^L_5 - {\bf v}_5 \|$ and $\Delta V_6 = \| {\bf v}^H - {\bf v}^L_6 \|$. Section \ref{sec_flyby} presents an analytical method to select the flyby parameters yielding the optimal ${\bf v}_5$. For the time being, it will be assumed that ${\bf v}_5$ is known.

For the Jupiter-Saturn leg ${\bf r}_3={\bf r}^J$, ${\bf r}_4 = {\bf r}^S$ and the duration is $t_4-t_3$. Only the first impulse $\Delta V_3 = \| {\bf v}^L_3 - {\bf v}_3 \|$ of the segment is relevant, because there is no rendezvous with Saturn.

\subsubsection{Determination of the optimal flyby parameters}
\label{sec_flyby}
The discussion below uses the Jupiter flyby as example. The procedure is identical for the GA with Saturn.

\begin{figure}
	\centering
	\includegraphics[width=0.40\textwidth]{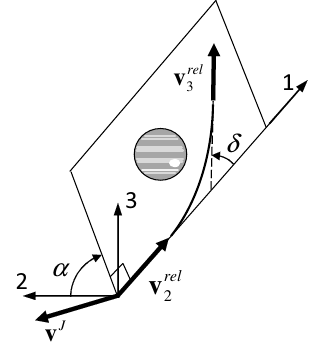}
	\caption{Flyby geometry in planetocentric frame $\{ {\bf u}_1 , {\bf u}_2 , {\bf u}_3 \}$.}
	\label{fig:PlanetRefFrame}
\end{figure}

For a given $\{C3, t_5\}$, the Jupiter arrival velocity ${\bf v}_2$ and the Lambert terminal velocity ${\bf v}^L_3$ are known. The objective is to find the combination of deflection angle $\delta \in ]0,\pi[$ of the relative velocity, and inclination of the planetocentric hyperbola $\alpha \in [0,2\pi[$ that minimizes $\Delta V_3$. The deflection is related to the pericenter radius ($r_\pi$) through \citep{Broucke:1988}
\begin{equation} \label{eq_deltamax}
	\sin\left( \frac{\delta}{2} \right) = \frac{1} {1 + \frac{r_{\pi} v^2_\infty}{\mu^J}},
\end{equation}
where $v_\infty$ denotes the magnitude of the inbound relative velocity ${\bf v}^{rel}_2 = {\bf v}_2 - {\bf v}^J$ and $\mu^J$ is the gravitational parameter of Jupiter. The post-GA velocity is obtained from ${\bf v}_3 = {\bf v}^{rel}_3 + {\bf v}^J$. The angle between ${\bf v}^{rel}_2$ and ${\bf v}^{rel}_3$ is $\delta$ (Fig.~\ref{fig:PlanetRefFrame}), and have the same magnitude. 

Let us define a planetocentric reference frame with directions $\{ {\bf u}_1 , {\bf u}_2 , {\bf u}_3 \}$. The first direction is taken parallel to the inbound relative velocity
\begin{equation} \label{eq_u1}
	{\bf u}_1 = \frac {{\bf v}^{rel}_2} {v_\infty}.
\end{equation}
The third direction is orthogonal to the planet's velocity and ${\bf u}_1$:
\begin{equation} \label{eq_u3}
	{\bf u}_3 = \frac {{\bf u}_1 \times {\bf v}^J} {\| {\bf u}_1 \times {\bf v}^J \|}.
\end{equation}
Finally, ${\bf u}_2 = {\bf u}_3 \times {\bf u}_1$ completes the right-handed triad.
In this system, the components of ${\bf v}_3$ are
\begin{equation}
	(v_\infty\cos\delta + v^J_1 , v_\infty\sin\delta\cos\alpha  + v^J_2 , v_\infty\sin\delta\sin\alpha),
\end{equation}
where $v^J_i = {\bf v}^J \cdot {\bf u}_i$. Note that, by construction, $v^J_3 = 0$. Let ${\bf d} = {\bf v}^J - {\bf v}^L_3$. The components of ${\bf v}_3 - {\bf v}^L_3$ in the $123$ system are:
\begin{equation} \label{eq_v3mvl}
	(v_\infty\cos\delta + d_1 , v_\infty\sin\delta\cos\alpha  + d_2 , v_\infty\sin\delta\sin\alpha + d_3).
\end{equation}
The post-GA impulse magnitude can be computed from
\begin{equation} \label{eq_dv32a}
	\Delta V^2_3 = ({\bf v}_3 - {\bf v}^L_3) \cdot ({\bf v}_3 - {\bf v}^L_3).
\end{equation}
Inserting Eq.~\ref{eq_v3mvl} into Eq.~\ref{eq_dv32a} yields, after some tedious but straightforward algebra, 
\begin{equation} \label{eq_dv32b}
	\Delta V^2_3 = K + 2v_\infty (d_1\cos\delta + d_2\sin\delta\cos\alpha  + d_3\sin\delta\sin\alpha),
\end{equation}
with $K = v^2_\infty + d^2$. The flyby parameters $\{\delta$, $\alpha\}$ that minimize $\Delta V_3$ can be determined from the derivatives of Eq.~\ref{eq_dv32b}.
\begin{equation} \label{eq_der1del}
	\frac{1}{2v_\infty} \frac{\partial \Delta V^2_3}{\partial \delta} = \cos\delta (d_2\cos\alpha + d_3\sin\alpha) - d_1\sin\delta,
\end{equation}
\begin{equation} \label{eq_der1i}
	\frac{1}{2v_\infty} \frac{\partial \Delta V^2_3}{\partial \alpha} = \sin\delta (d_3\cos\alpha - d_2\sin\alpha),
\end{equation}
\begin{equation} \label{eq_der2del}
	\frac{1}{2v_\infty} \frac{\partial^2 \Delta V^2_3}{\partial \delta^2} = -\sin\delta (d_2\cos\alpha + d_3\sin\alpha) - d_1\cos\delta,
\end{equation}
\begin{equation} \label{eq_der2i}
	\frac{1}{2v_\infty} \frac{\partial^2 \Delta V^2_3}{\partial \alpha^2} = -\sin\delta (d_3\sin\alpha + d_2\cos\alpha).
\end{equation}
The optimal inclination is obtained setting Eq.~\ref{eq_der1i} to zero. This yields three possible solutions. The trivial one ($\delta=0$) is not relevant, because it corresponds to no gravitational interaction with the planet. The other two, spaced $\pi$ apart, are given by
\begin{equation} \label{eq_iopt}
	\tan\alpha = \frac{d_3}{d_2}.
\end{equation}
One solution corresponds to a local maximum of $\Delta V_3$ and must be discarded. The optimal inclination $\alpha_{opt}$ makes Eq.~\ref{eq_der2i} positive. Because $\delta \in ]0,\pi[$, $\sin\delta$ is positive, therefore:
\begin{equation} \label{eq_cond_alpmin}
	d_3\sin\alpha_{opt} + d_2\cos\alpha_{opt} < 0.
\end{equation}
Next, the optimal deflection $\delta_{opt}$ is computed from Eq.~\ref{eq_der1del}
\begin{equation} \label{eq_delopt}
	\tan\delta = \frac{d_2\cos\alpha_{opt} + d_3\sin\alpha_{opt}}{d_1}.
\end{equation}
Only one solution of Eq.~\ref{eq_delopt} is relevant, because $\delta \in ]0,\pi[$. If this solution is indeed a minimum, Eq.~\ref{eq_der2del} must be positive. Therefore:
\begin{equation} \label{eq_cond_delmin}
	d_2\cos\alpha_{opt} + d_3\sin\alpha_{opt} + d_1\cot\delta_{opt} < 0.
\end{equation}
Substituting Eq.~\ref{eq_delopt} into Eq.~\ref{eq_cond_delmin} gives the condition
\begin{equation} \label{eq_cond_delmin_2}
	\frac {d^2_1 + (d_2\cos\alpha_{opt} + d_3\sin\alpha_{opt})^2} {d_2\cos\alpha_{opt} + d_3\sin\alpha_{opt}} < 0.
\end{equation}
Because the numerator is always positive, and the denominator is negative by virtue of Eq.~\ref{eq_cond_alpmin}, the inequality is automatically true and the extremum is always a minimum of $\Delta V_3$.

The optimum solution is only feasible if $\delta_{opt} \leq \delta_{max}$, where $\delta_{max}$ is obtained inserting the minimum acceptable flyby pericenter radius in Eq.~\ref{eq_deltamax}. If $\delta_{opt}$ is too large, the absolute minimum must lie either at $0$ or $\delta_{max}$. In this case, use $\delta = \delta_{max}$. The minimum cannot be at $\delta = 0$, because it would require the existence of a local maximum between $0$ and $\delta_{opt}$.

The computation of the optimal flyby can be simplified using a geometric interpretation. Equation~\ref{eq_dvt} can be written as
\begin{equation} \label{eq_podv}
	\Delta V^2_3 = K + 2v_\infty {\bf e} \cdot {\bf d},
\end{equation}
where the components of ${\bf e}$ in the 123 system are $(\cos\delta, \sin\delta\cos\alpha, \sin\delta\sin\alpha)$. Because ${\bf d}$ is constant and $\| {\bf e} \| = 1$, the dot product is minimal when ${\bf e}$ and ${\bf d}$ are antiparallel. This is possible as long as $\delta$ is not constrained, because ${\bf e}$ can take any direction. Thus,
\begin{equation} \label{eq_poso}
	{\bf e}_{opt} = -\frac{\bf d}{\| {\bf d} \|} \quad \Rightarrow \quad \left\{ {\begin{array}{l}
			{\cos \delta_{opt}  = \frac{-d_1}{d}}\\
			{\sin \delta_{opt} \cos \alpha_{opt}  = \frac{-d_2}{d}} \quad , \\
			{\sin \delta_{opt} \sin \alpha_{opt}  = \frac{-d_3}{d}}
	\end{array}} \right.
\end{equation}
The optimal solution is
\begin{equation} \label{eq_pode}
	\delta_{opt} = \arccos \left( \frac{-d_1}{d} \right) \quad \text{with} \quad \delta_{opt} \in ]0,\pi[,
\end{equation}
\begin{equation} \label{eq_poal}
	\alpha_{opt} = \text{ATAN2} \left( -d_3, -d_2 \right),
\end{equation}
where ATAN2 denotes the partial arctangent function, such that ATAN2$(b,a)$ is the argument of the complex number $a+bi$. When $\delta$ is constrained by the flyby depth, the geometric derivation is not applicable. In this case, it has already been demonstrated that $\alpha_{opt}$ from Eq.~\ref{eq_poal} can be retained, while $\delta_{opt}$ must be replaced with $\delta_{max}$.

The procedure outlined above determines the optimal GA geometry directly from the arrival conditions to the planet and the first terminal velocity of the subsequent Lambert arc.

\subsubsection{Problem dimensionality and total mission cost}
\label{prob_dim_cost}
For a given launch year $y_1$, the Jupiter arrival time and state depend on $C3$ only, as shown in Sec.~\ref{sec_seg_12}. The position of Saturn during the corresponding flyby is a function of $t_5$ alone. Thus, the Lambert terminal velocities $\{ {\bf v}^L_3, {\bf v}^L_4 \}$ are determined completely by $\{C3, t_5\}$. Using the optimal flyby expressions from Sec.~\ref{sec_flyby}, the impulse $\Delta V_3$ can then be computed. For the last leg, the terminal velocities $\{ {\bf v}^L_5, {\bf v}^L_6 \}$ are functions of $\{t_5, t_6\}$ alone. Because ${\bf v}_4={\bf v}^L_4$ is known from the previous leg, the optimal GA geometry yields $\Delta V_5$, Finally, the rendezvous impulse is obtained from $\Delta V_6 = \| {\bf v}^H - {\bf v}^L_6 \|$. Therefore, the complete mission profile can be determined from the triad $\{C3, t_5, t_6\}$ and $y_1$.

The total mission cost using impulsive maneuvers in the two last legs is given by 
\begin{equation} \label{eq_dvt}
	\Delta V_{\textrm{imp}} = \Delta V_{EJ} + \Delta V_3 + \Delta V_5 + \Delta V_6,
\end{equation}
where $\Delta V_{EJ}$ denotes the impulse delivered by the electric propulsion system during the Earth-Jupiter leg
\begin{equation} \label{eq_dv12}
	\Delta V_{EJ} = I_{sp} g_0 \ln \left( \frac{m_1}{m_2} \right).
\end{equation}

\subsection{Low-thrust transcription scheme}
\label{sec_ltt}
The transcription strategy finds a thrust law that connects two points with a given ToF, with fixed initial velocity and mass. If the leg ends in a rendezvous maneuver, the final velocity is also prescribed. This way, the impulsive maneuvers can be replaced with LT arcs without affecting the dimensionality of the problem. Moreover, the calculation is mostly explicit, only the final mass requires an iterative solution. Therefore, the algorithm is very efficient computationally. It is worth pointing out that the method does not guarantee the optimality of the control law, but yields feasible trajectories. This is appropriate for the scope of the study, which focuses on the viability of the mission concept.

The scheme will be illustrated with the Saturn-Halley leg. The trajectory is subdivided into $N$ regular time intervals, such that the time increment is $\Delta t = (t_6 - t_5)/N$. The states at the endpoints of the sub-intervals are denoted as ${\bf x}_i = [{\bf r}_i, {\bf v}_i, m_i]^T$, with $i=0,1,...,N$. The Saturn departure state (${\bf x}_0$) is fully determined from the previous legs, while at rendezvous with the comet ${\bf r}_N$ and ${\bf v}_N$ are known but $m_N$ is not. A reasonable guess of the final mass $\tilde{m}_N$ is required, which shall be improved iteratively. This initial approximation can be obtained using the impulsive solution from Sec.~\ref{sec_seg_3456} and the rocket equation:
\begin{equation}
	\frac{m_0}{\tilde{m}_N} = \exp \left( \frac{\Delta V_5 + \Delta V_6}{I_{sp} g_0} \right).
\end{equation} 

\begin{figure}
	\centering
	\includegraphics[width=0.70\textwidth]{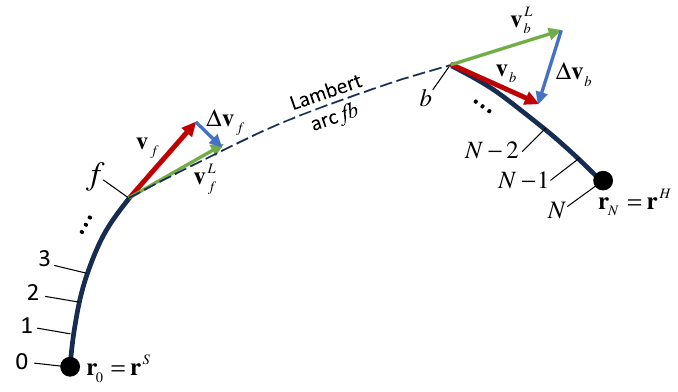}
	\caption{Low-thrust transcription scheme.}
	\label{fig:lt_scheme}
\end{figure}

The algorithm propagates the trajectory forward and backward in time from the two known end states. At each step, the thrust direction is chosen parallel to the residual velocities of the Lambert arc connecting the two working points. Let $f$ and $b$ denote the indices of the forward and backward propagation endpoints, respectively (Fig.~\ref{fig:lt_scheme}).

The procedure is initialized setting $f=0$ and $b=N$. Let ${\bf v}^L_f$ and ${\bf v}^L_b$ be the terminal velocities of the Lambert arc connecting ${\bf r}_f$ and ${\bf r}_b$, with a flight time $\Delta t (b-f)$. The residual velocities are given by
\begin{equation}
	\Delta {\bf v}_f = {\bf v}^L_f - {\bf v}_f \quad , \quad \Delta {\bf v}_b = {\bf v}_b - {\bf v}^L_b. 
\end{equation} 
The thrust directions are taken parallel to the residual velocities
\begin{equation}
	{\bf t}_i = \frac{ \Delta {\bf v}^L_i }{\| \Delta {\bf v}^L_i \|} \quad i=f,b. 
\end{equation} 
The thrust vectors are given by
\begin{equation} \label{eq_thrust}
	{\bf T}_i = \phi_i T_{max} {\bf t}_i \quad i=f,b,
\end{equation} 
where $\phi_i$ denotes the thrust setting
\begin{equation} \label{eq_thrset}
	\phi_i = \min \left( 1, \frac{\| \Delta {\bf v}_i \|}{I^{max}_i} \right) \quad i=f,b.
\end{equation} 
In Eq.~\ref{eq_thrset}, $I^{max}_i$ represents the maximum impulse that the propulsion system can deliver during a time step
\begin{equation}
	I^{max}_i = \frac{ T_{max} }{ m_i} \Delta t. 
\end{equation} 
The control parameter $\phi_i$ prevents the algorithm from overshooting the solution once the residual velocities become small. It ensures that, during one time step, the thruster does not deliver more impulse than what is required to cancel the residual velocity. 

Using the thrust vectors from Eq.~\ref{eq_thrust}, the trajectory is propagated numerically one step on both sides. That is, from $f$ to $f+1$ forward and from $b$ to $b-1$ backward in time. The process is repeated until both branches connect ($b=f+1$). If the trajectory is feasible, the final residual velocities are extremely small, typically below $10^{-10} \mathrm{~km~s^{-1}}$.

Because the backward propagation branch uses an estimate of the arrival mass (the dry mass of the spacecraft), there will be a mass mismatch between both branches. The initial dry mass estimate is refined iteratively with the secant method, until the mismatch falls below 1~g.

While the algorithm described above is functional, it suffers from a limitation. In most cases, one of the residual velocities decreases faster than the other. This causes one of the branches to operate at partial throttling ($\phi_i<1$) for substantial periods of time, reducing the total impulse that the propulsion system can deliver. This issue is easy to overcome, by advancing the two branches simultaneously only when both operate at maximum thrust ($\phi_f=\phi_b=1$). Otherwise, only the branch with the largest $\phi_i$ is advanced. That is:
\begin{itemize}
	\item Propagate from $f$ to $f+1$ if $\phi_f=1$ or $\phi_f>\phi_b$
	\item Propagate from $b$ to $b-1$ if $\phi_b=1$ or $\phi_b>\phi_f$
\end{itemize}

The transcription algorithm bears resemblance to the methodology presented by Beolchi et al. \cite{Beolchi:2026}. However, the reference modulates the time increments instead of the thrust magnitude.

For the Jupiter-Saturn segment, the final velocity is free. The procedure becomes much simpler because only the forward propagation branch is used. Therefore, $b$ is fixed ($b \equiv N$) and mass iterations are not required.

\section{Results and discussion}
\label{sec_res_dis}
The analysis begins with a comprehensive exploration of the parameter space, using impulsive trajectories for the Jupiter-Saturn-Halley segment. The simplified method enables a quick solution, and also reveals the structure of the complete $\{C3, t_5, t_6\}$ domain. If continuous thrust arcs were used, regions corresponding to trajectories that exceed the capabilities of the propulsion system could not be plotted.

\subsection{Preliminary exploration of the solution space}
\label{sec_prelim_expl}
The search for candidate solutions considers launch dates between 2030 and 2040, assuming an initial mass of 1500~kg. The minimum Jupiter pericenter radius is set at 100\,000~km, which is 25\,000~km higher that the orbit insertion burn of Juno \citep{Lam:2008}. For Saturn, the minimum is 80\,000~km, 1650~km higher than the insertion burn of Cassini-Huygens \citep{Matson:2002}.

The rendezvous date is constrained to the interval between the comet's crossing of the orbits of Uranus (June 2053) and Mars (May 2061). Trajectories that intercept the comet farther from the Sun than Uranus are too short, in the $ToF$ sense, for the propulsion system to provide the necessary impulse. Reaching the comet outside the orbit of Mars is a scientific requirement. It ensures that the beginning of the high-activity phase can be observed.

\begin{figure}
	\centering
	\includegraphics[width=0.70\textwidth]{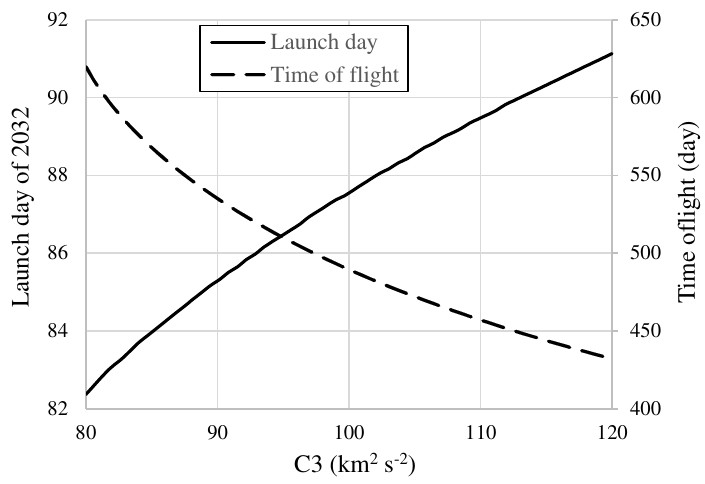}
	\caption{Launch day of 2032 (solid line) and Earth-Jupiter time of flight (dashed line) vs. $C3$.}
	\label{fig:2032_t1_t12}
\end{figure}

\begin{figure}
	\centering
	\includegraphics[width=0.70\textwidth]{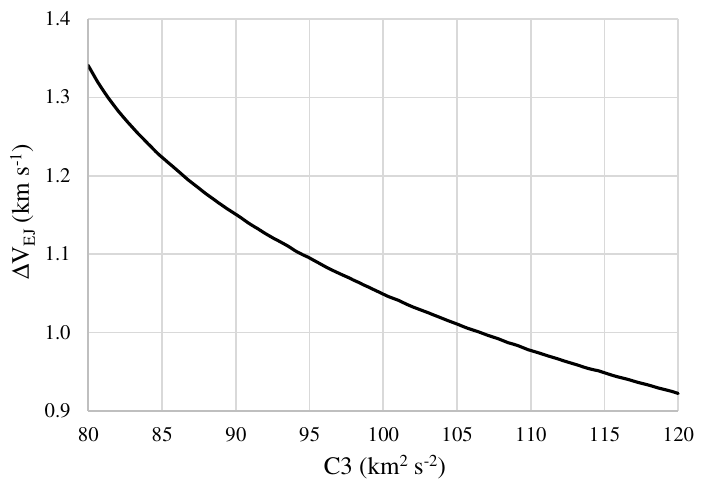}
	\caption{Earth-Jupiter leg impulse vs. $C3$ for launches in 2032.}
	\label{fig:2032_dvej}
\end{figure}

The Earth-Jupiter segment, because the orbits of the planets are close to circular and coplanar, displays the same general trends irrespective of the launch year. As an example, Fig.~\ref{fig:2032_t1_t12} displays the time of flight and launch day for the year 2032. The launches take place near the end of March, with the date varying by 9 days over the range of $C3$ considered. As $C3$ increases, the transfer time and angle both decrease, requiring a later launch to maintain the correct Earth-Jupiter phasing. Due to the synodic period of the two planets, the launch date moves forward by approximately 33 days every year (i.e., launches in 2033 take place by the end of April, and so on). Because the thrust and specific impulse are constant, $\Delta V_{EJ}$ is a function of the leg duration only. Its variation with $C3$ is shown in Fig.~\ref{fig:2032_dvej}. Due to the eccentricity of the planetary orbits, $\Delta V_{EJ}$ can experience variations on the order of $\mathrm{50~m~s^{-1}}$ from one year to another for a fixed $C3$.

\begin{figure}
	\centering
	\includegraphics[width=0.70\textwidth]{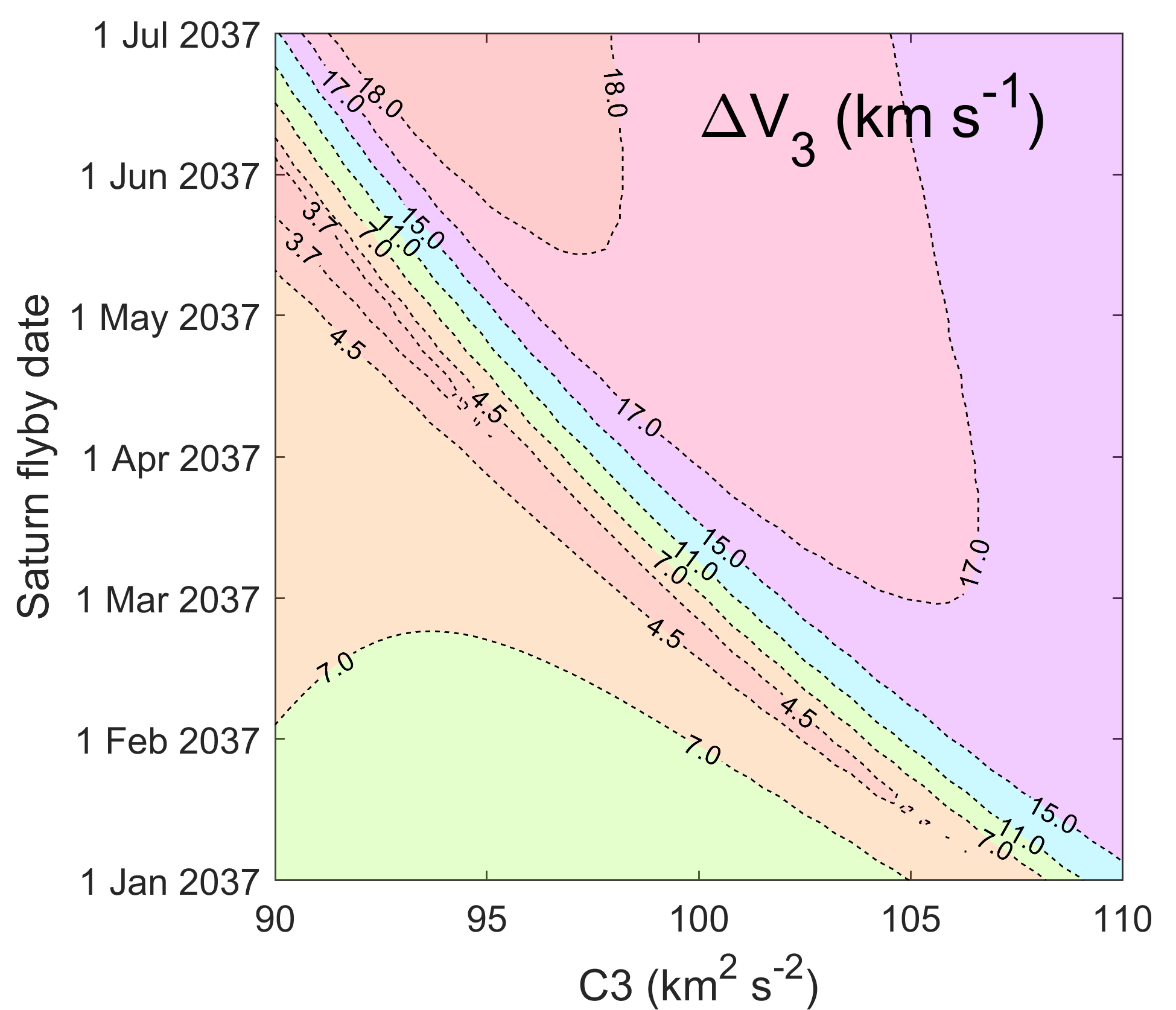}
	\caption{Impulse $\Delta V_3$ vs. $C3$ and Saturn flyby date. Launch in 2032.}
	\label{fig:2032_dv3_map}
\end{figure}

\begin{figure}
	\centering
	\includegraphics[width=0.70\textwidth]{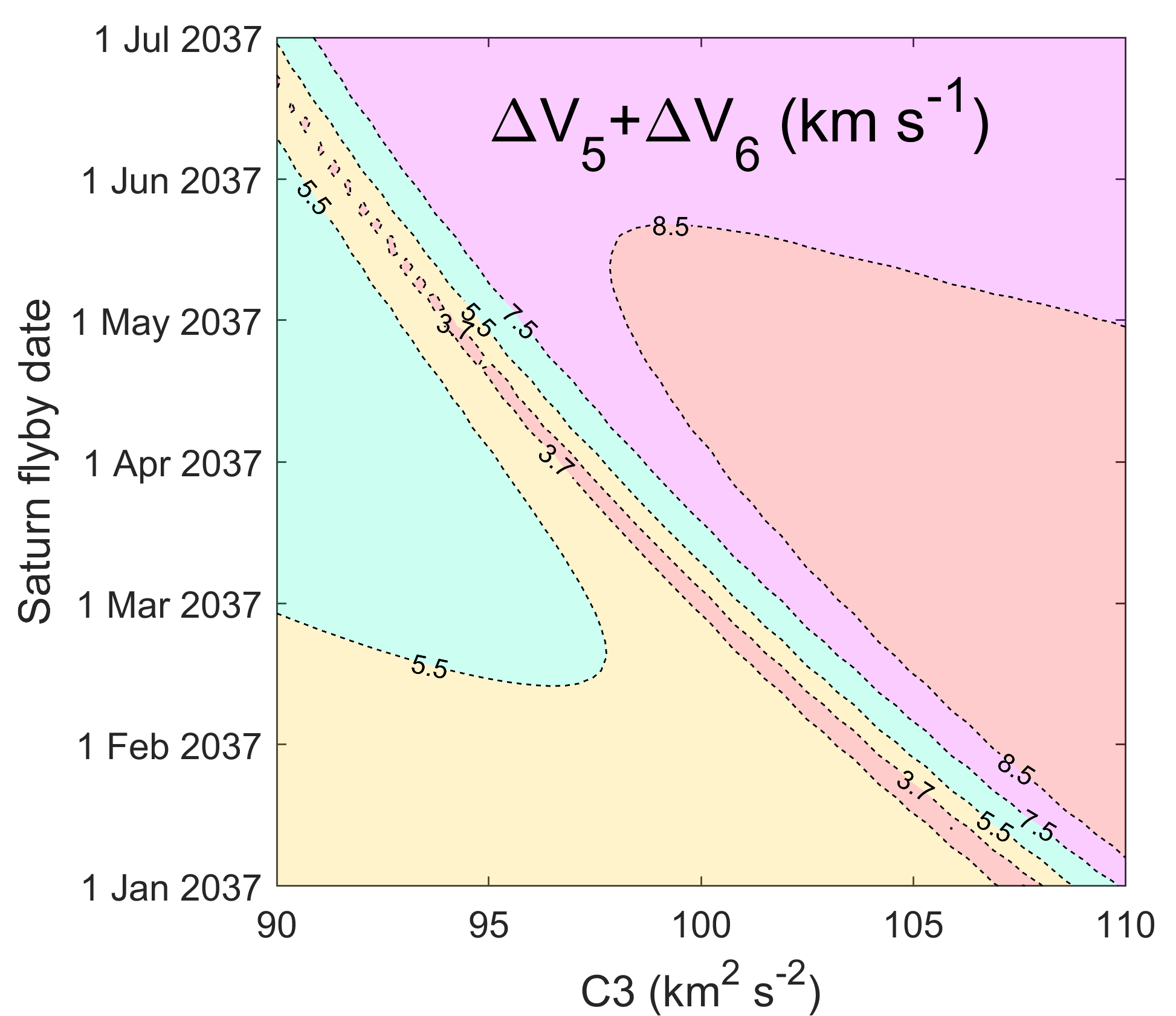}
	\caption{Last leg impulse vs. $C3$ and Saturn flyby date, for the optimal comet rendezvous epoch. Launch in 2032.}
	\label{fig:2032_dv56_map}
\end{figure}

The Jupiter-Saturn segment turns out to be critical. To achieve an acceptable $\Delta V_{\textrm{imp}}$, a high Jupiter departure speed, well above the escape velocity of the solar system, is required to increase the effectiveness of the Saturn GA. This makes the Jupiter-Saturn $ToF$ relatively short, between 1 and 4 years, depending on the phase of the planets. Under these conditions, the ability of the LT propulsion system to alter the trajectory is limited. Only solutions with moderate $\Delta V_3$ impulse, below $\mathrm{2~km~s^{-1}}$ for the longest Jupiter-Saturn transfers, are compatible with electric propulsion. For this reason, most of the launch dates between 2030 and 2040 are unfeasible. As an example, Fig.~\ref{fig:2032_dv3_map} plots $\Delta V_3$, as a function of launch energy and Saturn flyby date, for departures in 2032. The minimum impulse is close to $\mathrm{3.7~km~s^{-1}}$, exceeding the capability of the electric thruster.
This is unfortunate, because the Saturn phasing is favorable for the transfer to Halley. Figure~\ref{fig:2032_dv56_map} displays the total impulse of the last leg ($\Delta V_5 + \Delta V_6$) for a 2032 launch. Because the solution depends on 3 parameters, the 2D plot shows, for each $\{C3, t_5\}$ combination, the minimum impulse achievable over the acceptable range of $t_6$. That is, the impulse is minimized with respect to the rendezvous epoch. The minimum impulse for the last leg is modest, $\mathrm{3.7~km~s^{-1}}$. The rendezvous with the comet occurs between late 2058 and early 2060, depending on $C3$ and $t_5$, so the leg duration is above 20 years. Over this time span, the thruster would be fully capable of delivering the required impulse. However, due to the unfeasibility of the second leg, it is not possible to take advantage of the favorable Saturn-Halley alignment.

The analysis by  Beolchi et al. \cite{Beolchi:2024} showed that the impulse of the Saturn-Halley leg is minimized when the position of Saturn during the GA is close to the major axis of the comet's orbit. Roughly speaking, Saturn should be inside the projection of Halley's trajectory over the ecliptic. This happens between 2033 and 2038, approximately. Unfortunately, these Saturn flyby dates are incompatible with the limited impulse available for the Jupiter-Saturn leg.

Within the 2030-2040 time frame, the Jupiter and Saturn phases yield satisfactory solutions only for departures in 2036 and 2037. The major features of these trajectories are discussed below.

\subsubsection{Impulsive solution with launch in 2036}
\label{sec_2036_imp}
\begin{figure}
	\centering
	\includegraphics[width=0.70\textwidth]{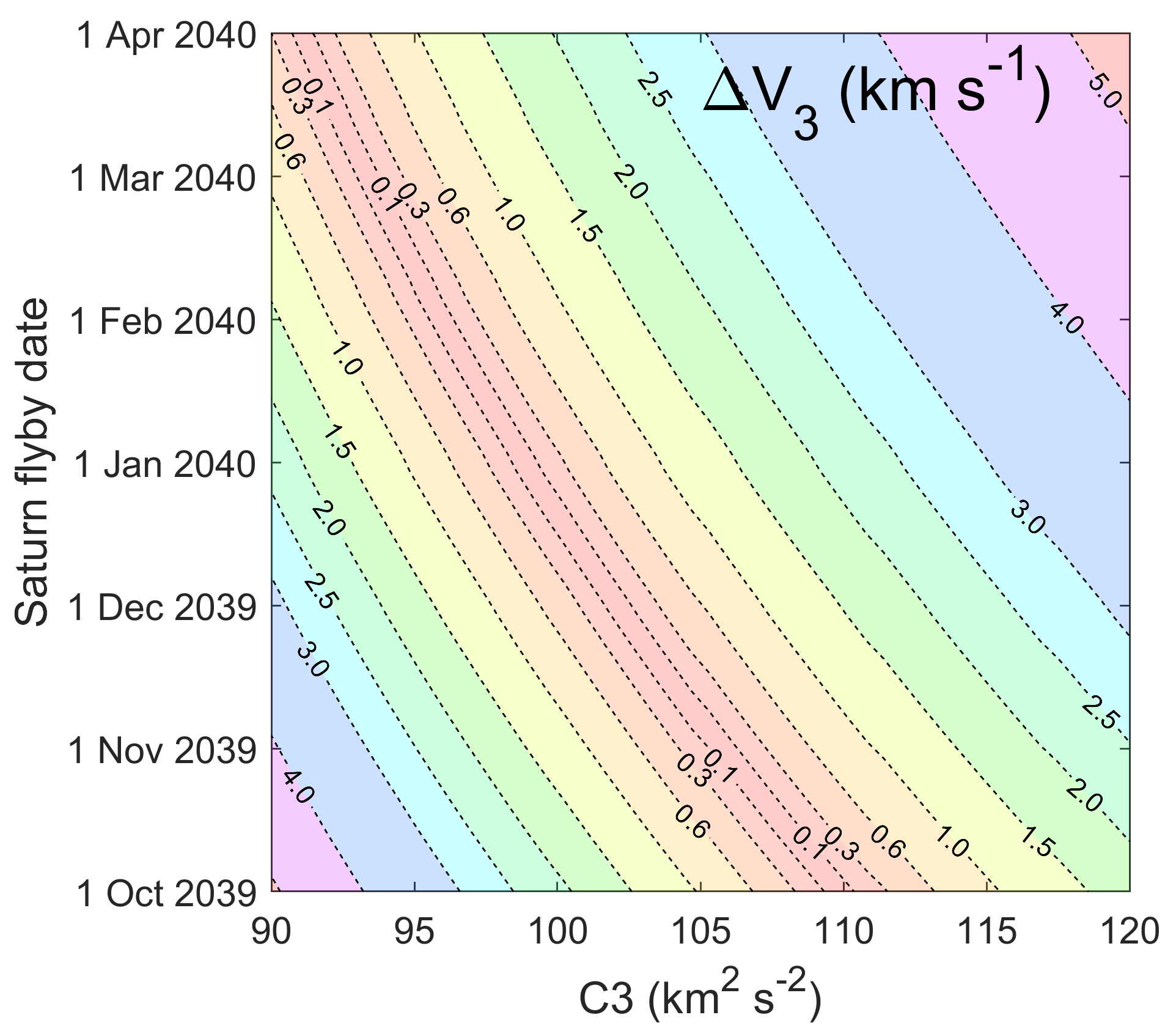}
	\caption{Impulse $\Delta V_3$ vs. $C3$ and Saturn flyby date. Launch in 2036.}
	\label{fig:2036_dv3_map}
\end{figure}

\begin{figure}
	\centering
	\includegraphics[width=0.70\textwidth]{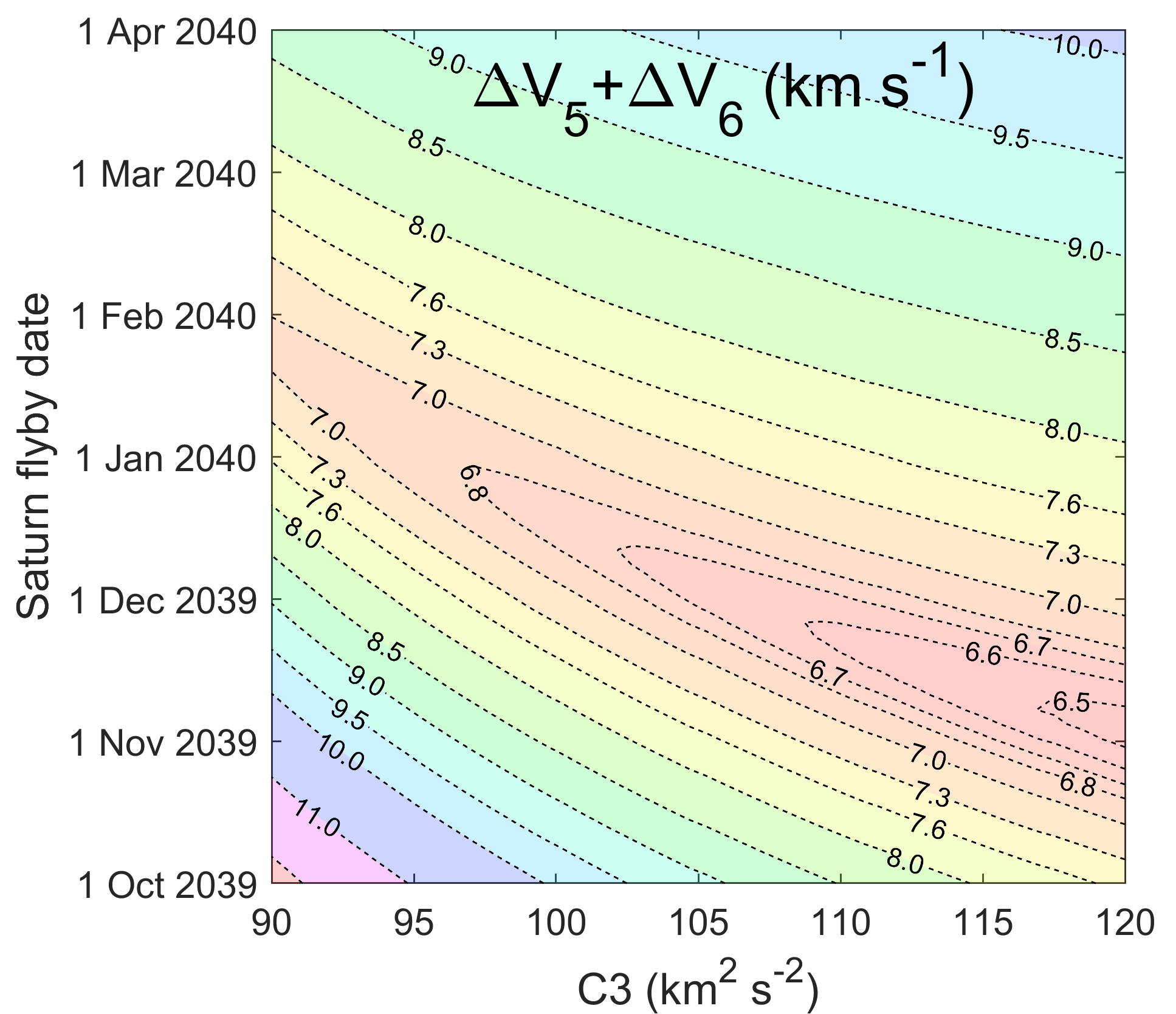}
	\caption{Last leg impulse vs. $C3$ and Saturn flyby date, for the optimal comet rendezvous epoch. Launch in 2036.}
	\label{fig:2036_dv56_map}
\end{figure}

As shown in Fig.~\ref{fig:2036_dv3_map}, there are $\{C3, t_5\}$ combinations resulting in a near-ballistic Jupiter-Saturn transfer ($\Delta V_3 < \mathrm{100~m~s^{-1}}$). However, because the Saturn GA takes place between late 2039 and early 2040, when the planet is far from the major axis of the comet's orbit, the last-leg impulse is above $\mathrm{6.5~km~s^{-1}}$ (Fig.~\ref{fig:2036_dv56_map}).

\begin{figure}
	\centering
	\includegraphics[width=0.70\textwidth]{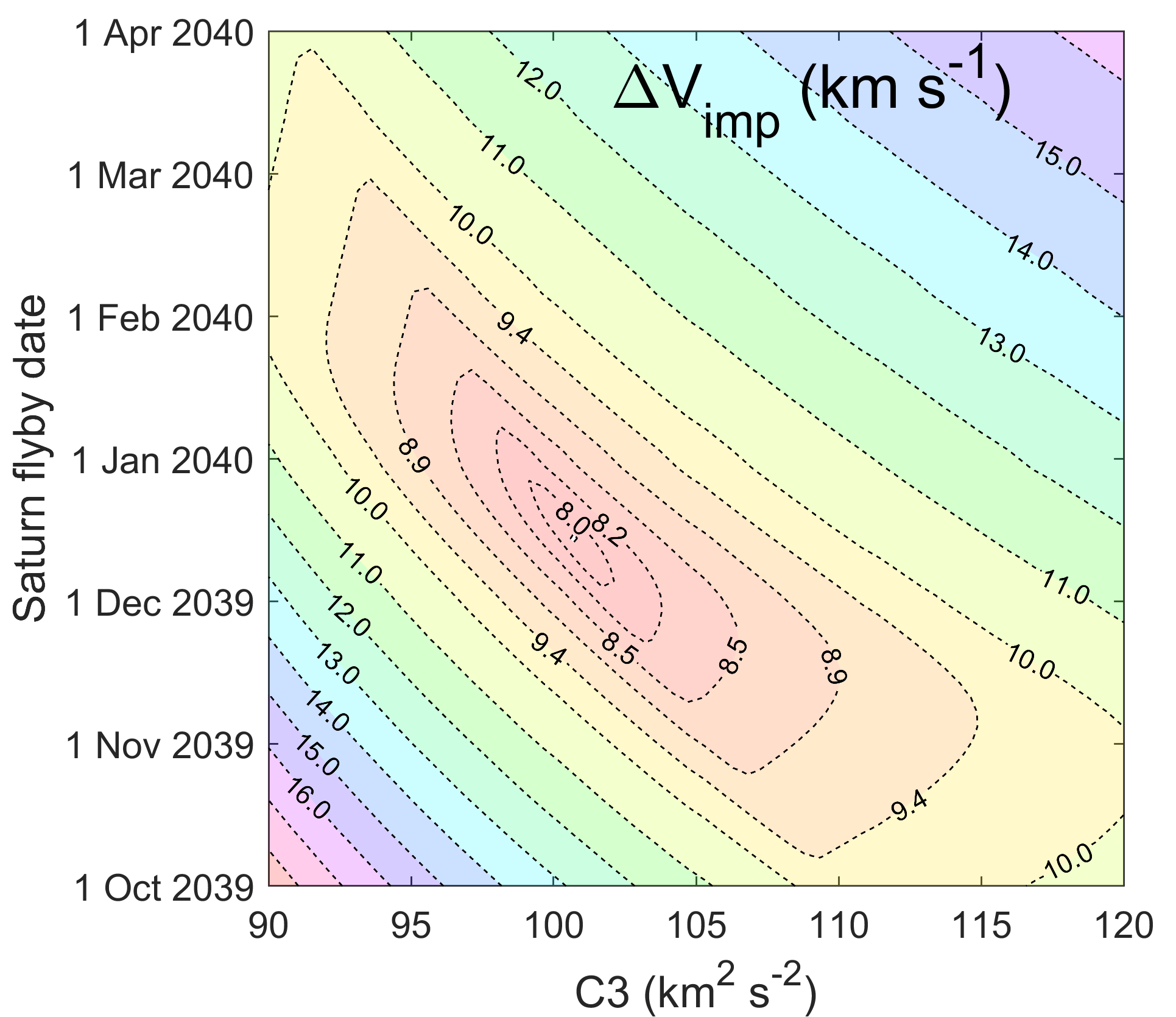}
	\caption{Total mission cost vs. $C3$ and Saturn flyby date, for the optimal comet rendezvous epoch. Launch in 2036.}
	\label{fig:2036_dvimp_map}
\end{figure}

The cost of the Earth-Jupiter transfer does not change substantially over the range of $C3$ considered (see Fig.~\ref{fig:2032_dvej}). Therefore, the optimal $\{C3, t_5, t_6\}$ combination is governed by the cost of the Jupiter-Halley transfer. Furthermore, $\Delta V_3$ grows rapidly when the Jupiter-Satrun leg deviates from a near-ballistic trajectory, to the point that it quickly dominates $\Delta V_{\textrm{imp}}$. Therefore, the optimal mission profile correspond to a point along the narrow quasi-ballistic band of Fig.~\ref{fig:2036_dv3_map}. Its intersection with the region of moderate last-leg impulse ($\sim \mathrm{6.7~km~s^{-1}}$) in Fig.~\ref{fig:2036_dv56_map} yields the minimum $\Delta V_{\textrm{imp}}$. The total mission cost is shown in Fig.~\ref{fig:2036_dvimp_map}, with a minimum value of $\mathrm{7.9~km~s^{-1}}$.

Table~\ref{tab_2036_imp} summarizes the optimal impulsive trajectory, which will be used as initial guess for the optimization with low-thrust arcs.

\begin{table}
	\centering
	\caption{Impulsive solution with launch in 2036.}
	\begin{tabular}{lclc} \hline
		Launch 					& 21 Aug 2036 	& $C3$ ($\mathrm{km^2s^{-2}}$) 	& 100 \\ \hline
		Jupiter GA 				& 4 Feb 2038  	& $\Delta V_{EJ}$ (km~s$^{-1}$)	& 1.14 \\ \hline
		Saturn GA 				& 21 Dec 2039 	& $\Delta V_3$ (km~s$^{-1}$) 	& 0.00 \\ \hline
		Rendezvous 				& 3 Mar 2060  	& $\Delta V_5$ (km~s$^{-1}$) 	& 0.63 \\ \hline
		$r_\pi^J$ ($10^3$ km)	& 149			& $\Delta V_6$ (km~s$^{-1}$) 	& 6.12 \\ \hline
		$r_\pi^S$ ($10^3$ km)	& 80			&								& \\ \hline
	\end{tabular}
	\label{tab_2036_imp}
\end{table} 

\subsubsection{Impulsive solution with launch in 2037}
\label{sec_2037_imp}
\begin{figure}
	\centering
	\includegraphics[width=0.70\textwidth]{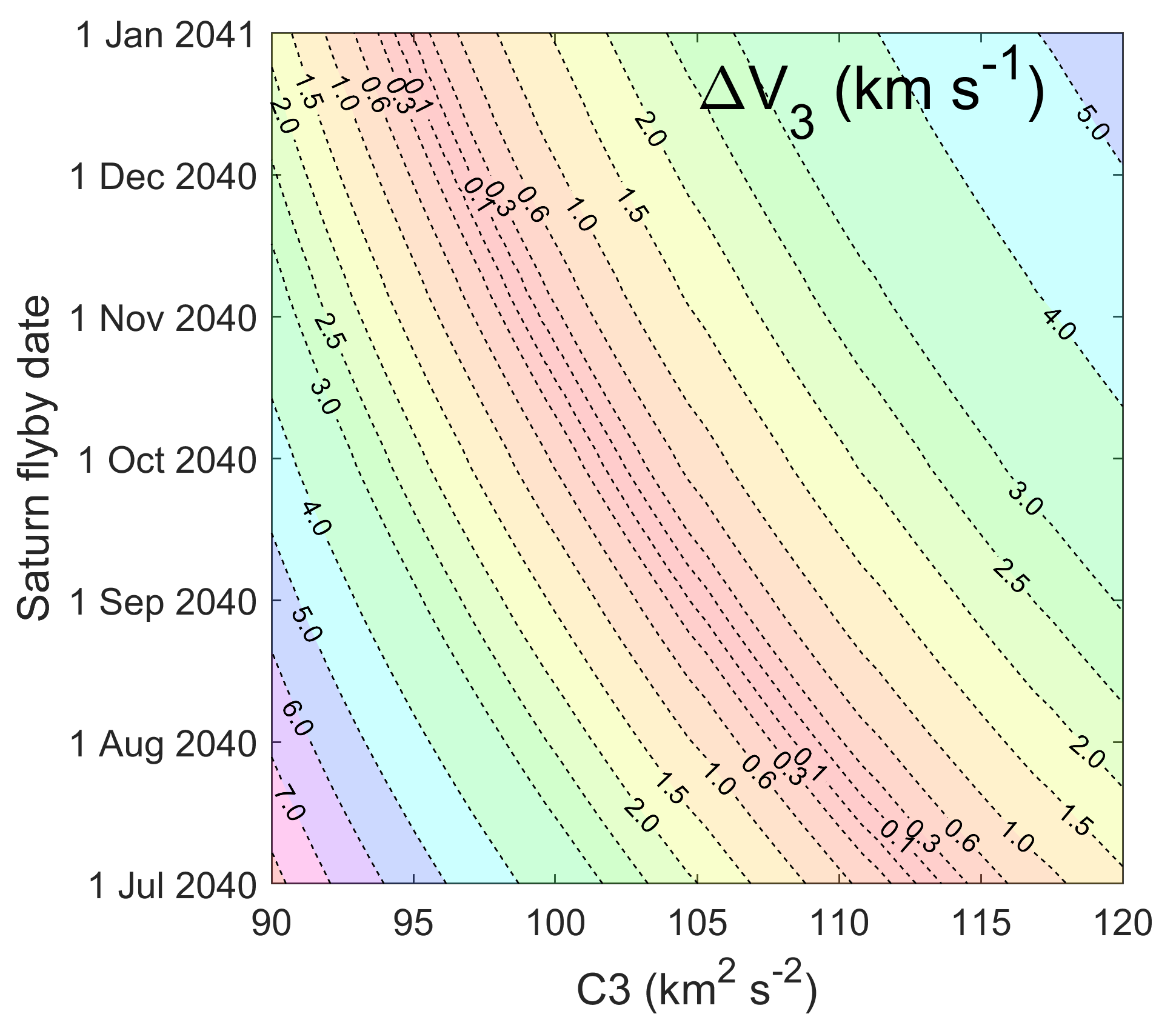}
	\caption{Impulse $\Delta V_3$ vs. $C3$ and Saturn flyby date. Launch in 2037.}
	\label{fig:2037_dv3_map}
\end{figure}

\begin{figure}
	\centering
	\includegraphics[width=0.70\textwidth]{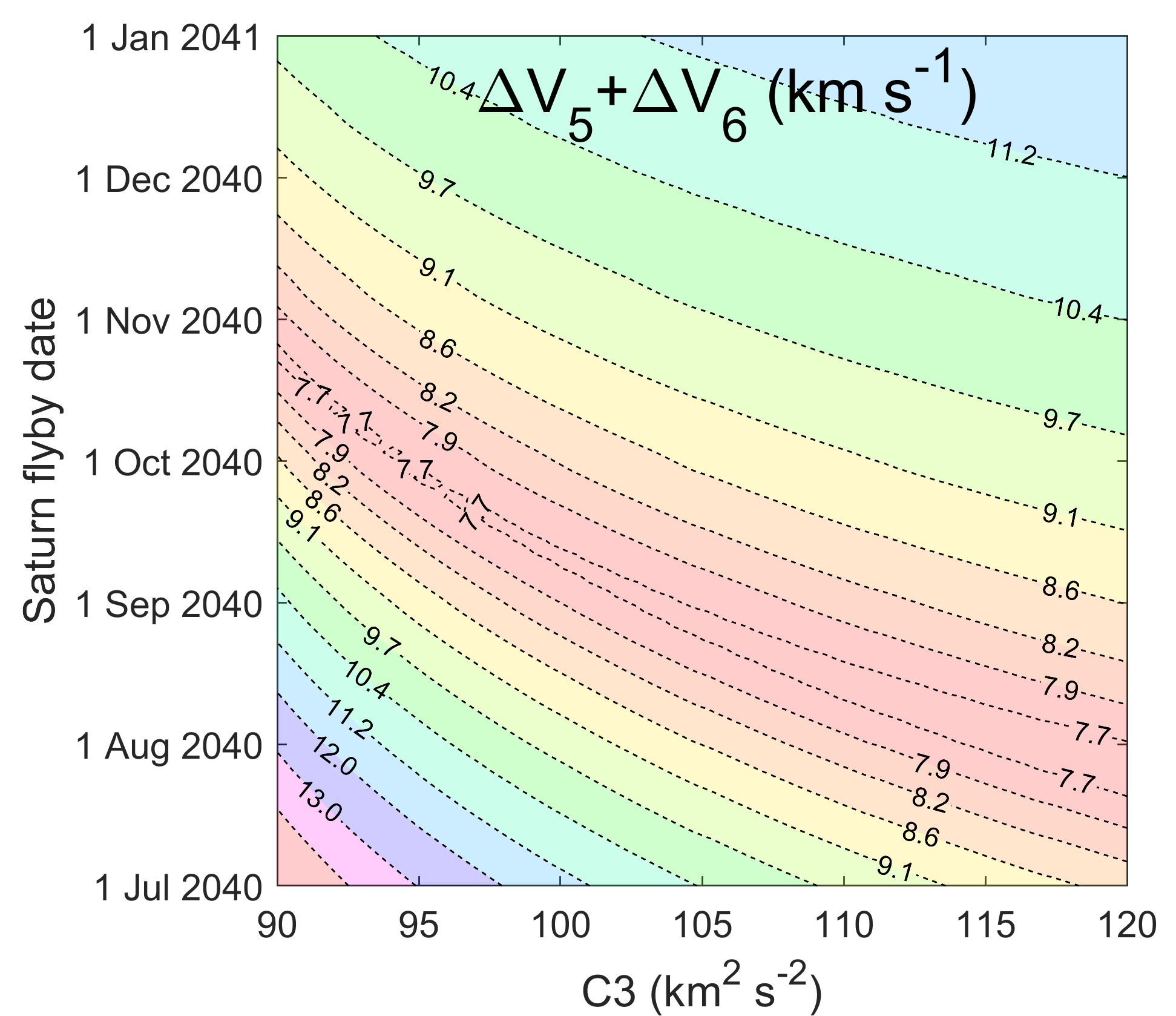}
	\caption{Last leg impulse vs. $C3$ and Saturn flyby date, for the optimal comet rendezvous epoch. Launch in 2037.}
	\label{fig:2037_dv56_map}
\end{figure}

\begin{figure}
	\centering
	\includegraphics[width=0.70\textwidth]{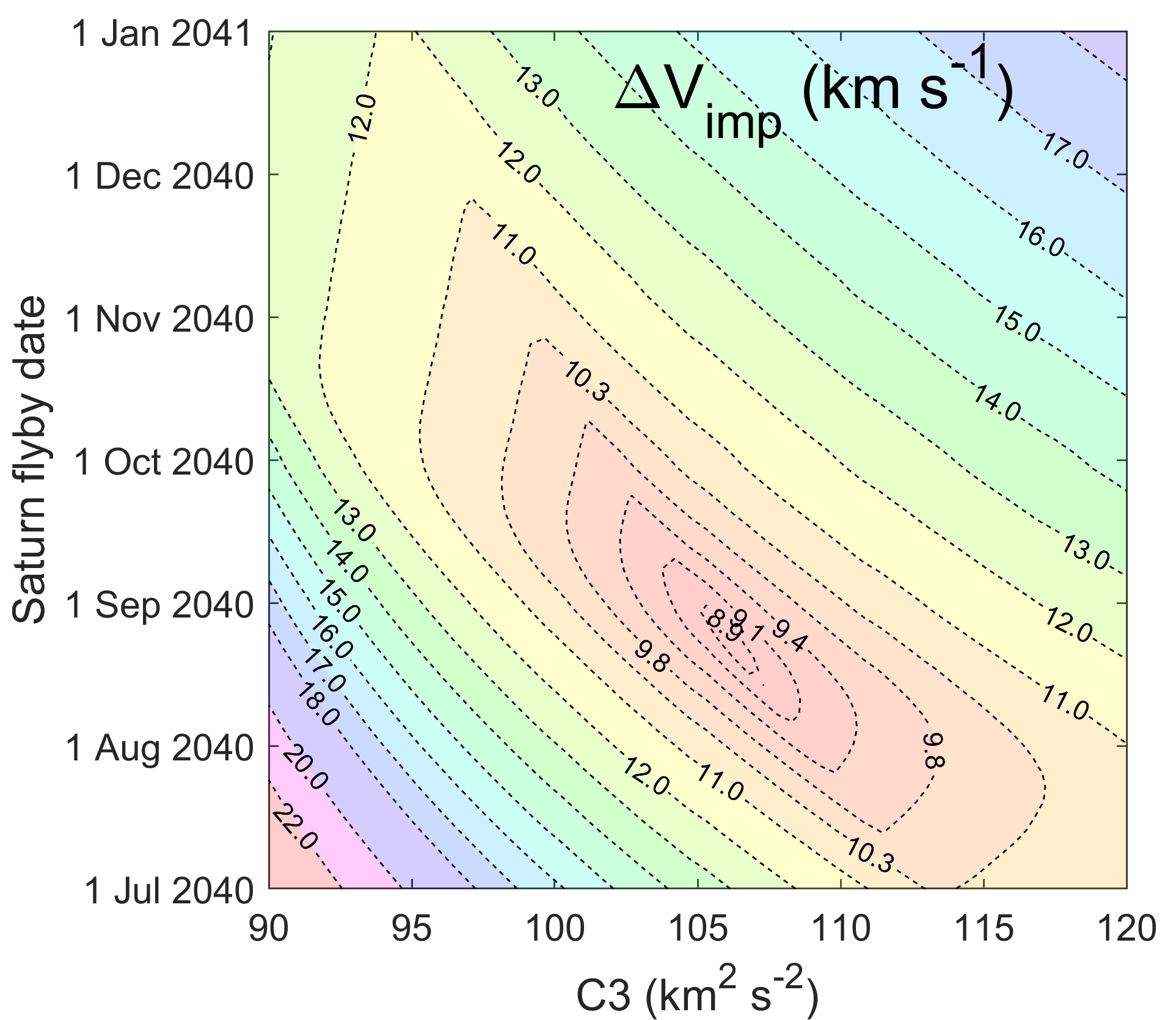}
	\caption{Total mission cost vs. $C3$ and Saturn flyby date, for the optimal comet rendezvous epoch. Launch in 2037.}
	\label{fig:2037_dvimp_map}
\end{figure}

The solution structure for launch in 2037 is very similar to the 2036 case. There is a narrow band in the $\{C3, t_5\}$ plane that results in quasi-ballistic Jupiter-Saturn transfers (Fig.~\ref{fig:2037_dv3_map}). Due to the later launch, Saturn is even farther from the optimal position during the flyby, resulting in last-leg impulses above $\mathrm{7.7~km~s^{-1}}$ (Fig.~\ref{fig:2037_dv56_map}). For this reason, the total impulse of the best solution (Table~\ref{tab_2037_imp}) is higher, $\mathrm{8.8~km~s^{-1}}$. See Fig.~\ref{fig:2037_dvimp_map} for the $\Delta V_{\textrm{imp}}$ map.

\begin{table}
	\centering
	\caption{Impulsive solution with launch in 2037.}
	\begin{tabular}{lclc} \hline
		Launch 					& 24 Sep 2037	& $C3$ ($\mathrm{km^2s^{-2}}$)	& 106 \\ \hline
		Jupiter GA 				& 28 Feb 2039	& $\Delta V_{EJ}$ (km~s$^{-1}$)	& 1.12 \\ \hline
		Saturn GA 				& 22 Aug 2040	& $\Delta V_3$ (km~s$^{-1}$)	& 0.00 \\ \hline
		Rendezvous 				& 15 Mar 2060	& $\Delta V_5$ (km~s$^{-1}$)	& 0.68 \\ \hline
		$r_\pi^J$ ($10^3$ km)	& 464			& $\Delta V_6$ (km~s$^{-1}$)	& 7.01 \\ \hline
		$r_\pi^S$ ($10^3$ km)	& 80			&								& \\ \hline
	\end{tabular}
	\label{tab_2037_imp}
\end{table} 

\subsection{Low-thrust trajectories}
\label{lt_trajec}
In this section, the $\{C3, t_5, t_6\}$ combinations yielding the minimum total impulse with LT arcs are sought. The solutions with impulsive maneuvers from Tables~\ref{tab_2036_imp} and \ref{tab_2037_imp} serve as initial guesses for the optimization. Each leg of the trajectory is discretized into 100 uniform time steps, propagating the spacecraft state with a RK4 integrator. For the Saturn-Jupiter-Halley segment, the LT transcription scheme from Sec.~\ref{sec_ltt} is used.

\subsubsection{Low-thrust solution with launch in 2036}
\label{sec_2036_lt}
Table~\ref{tab_2036_lt} summarizes the major events of the trajectory\footnote{Some values are given with 4 significant digits, for the sole purpose of highlighting the close similarity with the initial guess (Table~\ref{tab_2036_imp}), and evidencing that the Jupiter-Saturn leg is virtually ballistic. In reality, the simplicity of the dynamical model does not warrant this level of accuracy.}. The launch $C3$ is $\mathrm{99.96~km^2s^{-2}}$, almost identical to the impulsive initial guess (Table~\ref{tab_2036_imp}). This energy is easily achievable with any of the launchers in Fig.~\ref{fig:Launchers}. In fact, it would be possible to increase the initial mass at this $C3$ level. This is explored in Sec.~\ref{sec_var_mass}.

The launch and flyby dates are identical to the impulsive trajectory, demonstrating that it is an excellent initial guess for the continuous thrust solution. Only the rendezvous date changes substantially, occurring five months later. The encounter with the comet takes place 4.96~au from the Sun, just inside the orbit of Jupiter, giving ample margin to observe the onset of intense cometary activity.

Note that the Saturn flyby radius is the minimum allowed. A more efficient trajectory may be possible if a closer approach is allowed. The $r_\pi^S$ limit selected corresponds to an altitude 20\,000~km above the cloud top. There could be room for improvement in this aspect.

The total impulse with LT is $\mathrm{9.64~km~s^{-1}}$, vs. 8.81 with impulsive maneuvers. The higher $\Delta V$ is completely offset by the superior efficiency of electric propulsion. Assuming a chemical rocket with $I_{sp}=300$~s, the final mass of the impulsive solution would drop to 75~kg, which is completely unacceptable.
The Jupiter-Saturn impulse is only $\mathrm{2~m~s^{-1}}$. This is due to the extreme sensitivity of the second leg cost, which forces the solutions to become near-ballistic. The mass at rendezvous is 812~kg, comparable to the dry mass of Deep Impact (879~kg, including 364~kg of impactor payload \citep{Blume:2005}). As mentioned before, there is room to increase the spacecraft mass due to the moderate $C3$.

\begin{table}
	\centering
	\caption{Summary of LT solution with launch in 2036.}
	\begin{tabular}{lccc} \hline
		Event		& Date 			& Mass (kg)	& $\Delta V$ (km~s$^{-1}$) \\ \hline
		Launch		& 21 Aug 2036	& 1500		& 1.143 \\ \hline
		Jupiter GA	& 4 Feb 2038	& 1395		& 0.002 \\ \hline
		Saturn GA	& 21 Dec 2039	& 1394		& 8.494 \\ \hline
		Rendezvous	& 14 Aug 2060	& 812 		& \\ \hline
		$C3$ ($\mathrm{km^2s^{-2}}$)	& $r_\pi^J$ ($10^3$ km)	& $r_\pi^S$ ($10^3$ km)	& $r^H$ (au) \\ \hline
		$\quad 99.96$				& 149					& 80			& 4.96 \\ 
		\hline
	\end{tabular}
	\label{tab_2036_lt}
\end{table} 

\begin{figure}
	\centering
	\includegraphics[width=0.70\textwidth]{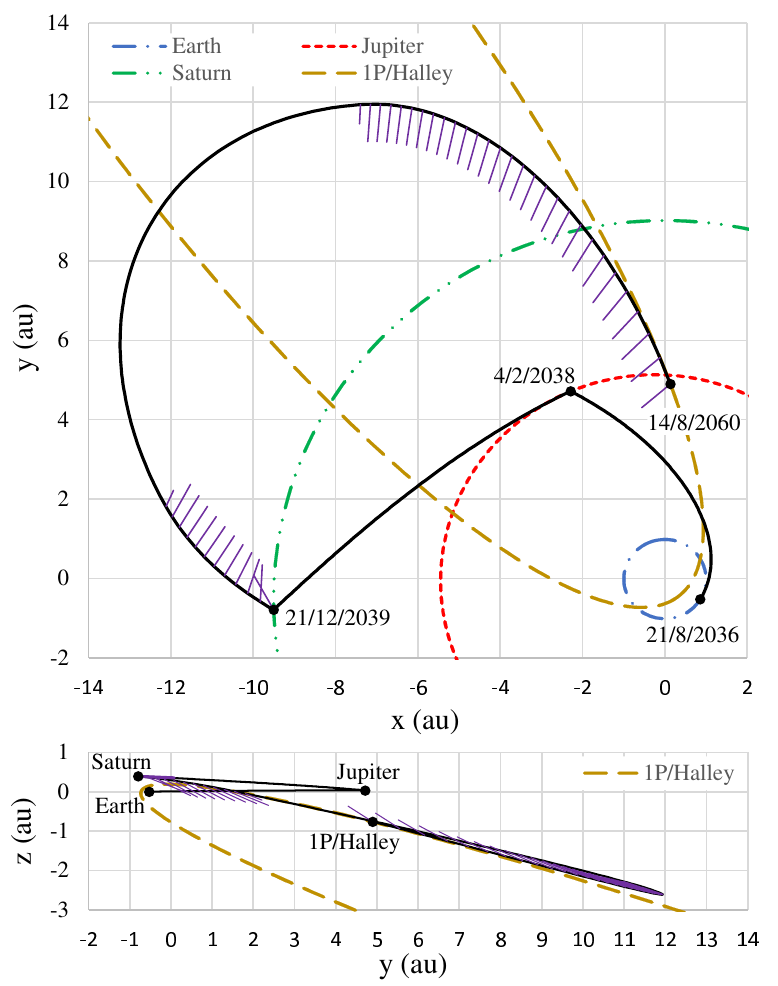}
	\caption{LT solution with 2036 launch. Continuous thick curve: trajectory, Short segments: thrust law, Dashed lines: planet and comet orbits.}
	\label{fig_2036_traj}
\end{figure}

Figure \ref{fig_2036_traj} shows the trajectory and the thrust law. The thrust lines are omitted for the Earth-Jupiter leg because the overlap with the trajectory (thrust is always parallel to the velocity vector).

The Jupiter GA provides a large boost to the spacecraft, the eccentricity of the osculating conic jumps from 0.813 to 2.558. That is, the probe is placed in a hyperbolic trajectory that, were it not to encounter Saturn, would eject it from the solar system. This ensures that the hyperbolic excess speed at Saturn is large enough ($\mathrm{16.3~km~s^{-1}}$) for a successful plane change. A previous mission concept with a single GA \citep{Beolchi:2024,Barbieri:2025} did not benefit from this energy boost. Instead, it compensated with a very high $C3$, up to $\mathrm{200~km^2s^{-2}}$ depending on the launch and flyby dates.

The Saturn GA places the spacecraft in a retrograde trajectory, draining a substantial part of its energy in the process. The osculating eccentricity drops to 0.807. The post-GA orbit is near-planar, indicating that the change of inclination is taken care of almost entirely by the Saturn flyby. The thrust acts mostly in-plane, curving the trajectory to adjust the eccentricity and accelerate the spacecraft until it matches the velocity of the comet.

\subsubsection{Low-thrust solution with launch in 2037}
\label{sec_2037_lt}
One synodic period (13 months) after the 2036 departure date, another launch window opens. Table~\ref{tab_2037_lt} presents the most relevant characteristics of this solution. Again, the impulsive approximation (Table~\ref{tab_2037_imp}) is an excellent initial guess for the LT impulse minimization. Compared to the $y_1=2036$ solution, the $C3$ requirement increases to $\mathrm{106~km^2s^{-2}}$. The total impulse is $\mathrm{1.23~km~s^{-1}}$ higher, due to the less desirable position of Saturn. The Jupiter-Saturn segment remains quasi-ballistic and the Saturn GA is also constrained by the minimum flyby altitude. The change in rendezvous date relative to the 2036 solution is limited, just 40 days. Therefore, the rendezvous occurs closer to the Sun, at 4.58~au, but still well outside the orbit of Mars.

\begin{figure}
	\centering
	\includegraphics[width=0.70\textwidth]{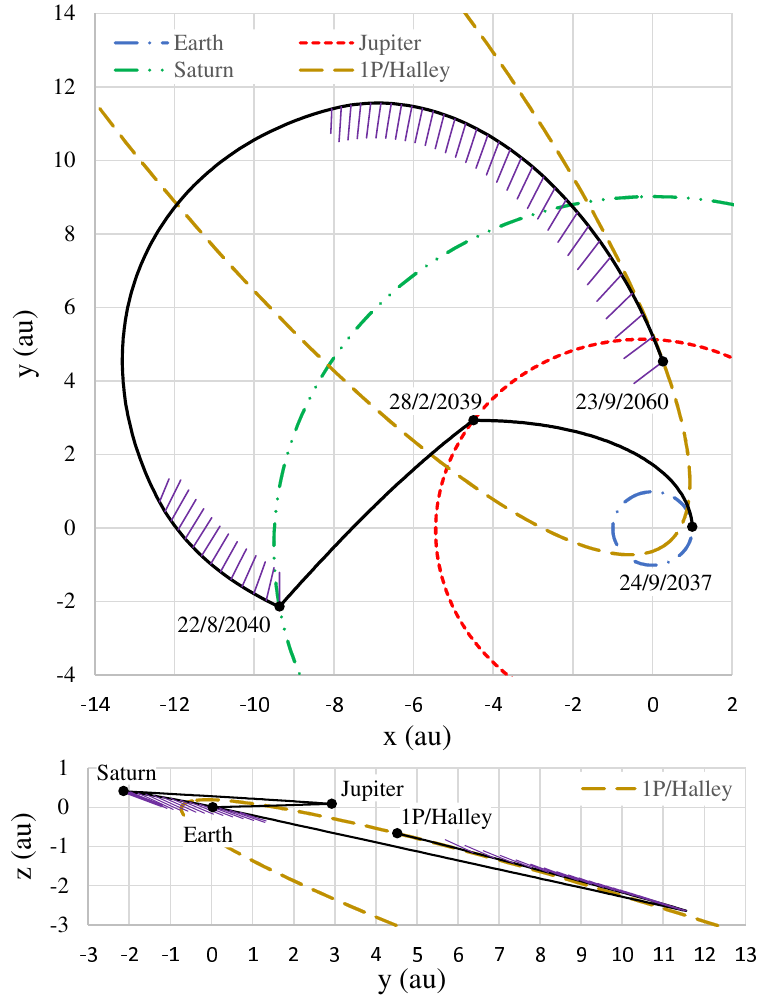}
	\caption{LT solution with 2037 launch. Continuous thick curve: trajectory, Short segments: thrust law, Dashed lines: planet and comet orbits.}
	\label{fig_2037_traj}
\end{figure}

The trajectory and thrust vectors are shown in Fig.~\ref{fig_2037_traj}. Other than the unfavorable position of Saturn, the structure of the trajectory is very similar to the $y_1=2036$ solution, and will not be discussed again. 

\begin{table}
	\centering
	\caption{Summary of LT solution with launch in 2037.}
	\begin{tabular}{lccc}
		\hline
		Event		& Date 			& Mass (kg)	& $\Delta V$ (km~s$^{-1}$) \\ \hline
		Launch		& 24 Sep 2037	& 1500		& 1.121 \\ \hline
		Jupiter GA	& 28 Feb 2039	& 1397		& 0.001 \\ \hline
		Saturn GA	& 22 Aug 2040	& 1397		& 9.743 \\ \hline
		Rendezvous	& 23 Sep 2060	& 751 		& \\ 
		\hline
		$C3$ ($\mathrm{km^2s^{-2}}$)	& $r_\pi^J$ ($10^3$ km)	& $r_\pi^S$ ($10^3$ km)	& $r^H$ (au) \\ \hline
		$\quad 106.28$					& 464					& 80			& 4.58 \\ 
		\hline
	\end{tabular}
	\label{tab_2037_lt}
\end{table} 

\subsection{Variation of the launch mass}
\label{sec_var_mass} 
\begin{figure}
	\centering
	\includegraphics[width=0.70\textwidth]{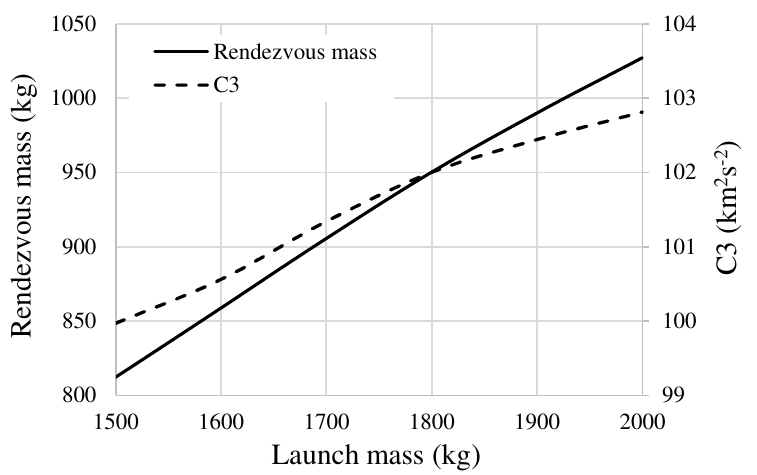}
	\caption{Final mass (solid line) and $C3$ (dashed line) vs. launch mass for departure in 2036.}
	\label{fig_2036_mass}
\end{figure}

The launchers in Fig.~\ref{fig:Launchers} have a payload capability above 1500 kg at the launch energies required by the LT solutions. Therefore, it is possible to increase the scientific payload by raising the launch mass without compromising the feasibility of the solution. Figure~\ref{fig_2036_mass} shows the variation of the dry mass and $C3$ as a function of the initial mass, for a 2036 launch.

A wet mass of two tonnes raises the rendezvous mass to 1027~kg at the cost increasing $C3$ to $\mathrm{102.8~km^2s^{-2}}$. A higher launch energy is required because the thrust of the motor does not scale with the mass, reducing its ability to increase the specific energy of the spacecraft. To reach Jupiter with the same velocity, a heavier spacecraft requires higher $C3$. The departure date is highly insensitive to the launch mass. The 2000~kg spacecraft launches on 22 August 2036, only one day after the 1500 kg probe. The rendezvous date experiences a larger variation, moving to 23 September 2060, a delay of 40 days. Consequently, the spacecraft reaches the comet closer to the Sun, at 4.57~au. 

\begin{figure}
	\centering
	\includegraphics[width=0.70\textwidth]{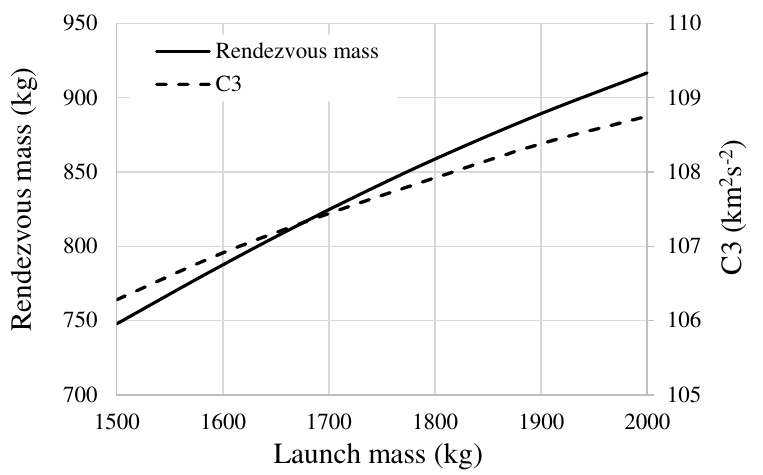}
	\caption{Final mass (solid line) and $C3$ (dashed line) vs. launch mass for departure in 2037.}
	\label{fig_2037_mass}
\end{figure}

The effect of varying the initial mass for a launch in 2037 is shown in Fig.~\ref{fig_2037_mass}. A two-tonne wet mass gives a dry mass of 917~kg, requirimg $C3= \mathrm{108.8~km^2s^{-2}}$. As was the case for the 2036 departure, the launch date changes only by one day when the mass is increased to 2000~kg. In contrast, the rendezvous occurs on 18 September 2060, 5 days earlier, intercepting the comet 4.63~au from the Sun. 

\section{Conclusions}
\label{sec_con} 
The feasibility of a rendezvous mission with Halley's comet during its next approach to the Sun has been demonstrated. The comet's retrograde high-inclination eccentric orbit poses a major challenge. Contrary to existing studies, this trajectory relies exclusively on tested technologies. Thrust is provided by a 36~mN Hall-effect motor powered by RTGs, and the spacecraft's wet mass is compatible with existing heavy launchers. The encounter is constrained to occur before the comet crosses the orbit of Mars, when the high-activity phase begins. Gravity assist (GA) maneuvers with Jupiter and Saturn reduce propellant consumption.

The complete solution space is explored initially using impulsive transfers between Jupiter, Saturn and the comet. The impulsive solutions with the lowest cost serve as initial guesses for the optimization of the low-thrust (LT) trajectory. The LT law for the transfer between Jupiter and Halley is determined with an explicit transcription scheme. It propagates the spacecraft state forward and backward in time from the endpoints of each leg. Thrust acts in the direction of the Lambert impulses required to connect the end states of the remaining arc. The LT transcription scheme preserves the leg duration, maintaining the same number of design parameters as the impulsive solution. This strategy does not guarantee the optimality of the control law, but it demonstrates the feasibility of the trajectory. The fact that more refined design methods could enable lower propellant budgets lends further credibility to the mission concept.

The Earth-Jupiter leg uses continuous thrust to lower the characteristic launch energy ($C3$). Thrust acts parallel to the velocity vector, to extract maximum mechanical power. The initial spacecraft velocity is as close to parallel to Earth's motion as the inclination of Jupiter's orbit allows. This maximizes utilization of the launcher's energy. Under these assumptions, the initial and final states of the leg depend only on $C3$.

The GA with Jupiter inserts the spacecraft into a high-energy hyperbolic trajectory to Saturn. Using explicit expressions for the optimal GA geometry, it is possible to minimize the post-flyby impulse required to target Saturn. The Jupiter-Saturn leg can be parameterized as a function of $C3$ and the Saturn flyby date alone.

A Saturn GA changes the inclination to reach the orbital plane of Halley. Finally, the propulsion system adjusts the trajectory to complete the rendezvous with the comet. Characterizing this segment requires only one additional parameter, the arrival date.

Under the aforementioned assumptions, $C3$, Saturn flyby date and rendezvous date completely characterize the trajectory. The low dimensionality of the design space enables fast and robust optimization. For the majority of launch dates between 2030 and 2040, the impulse required for the Jupiter-Saturn leg exceeds the performance of the electric motor. The positions of Jupiter and Saturn yield satisfactory solutions only for launches in 2036 and 2037. 

The best solution with a 2036 departure, assuming a wet mass of 2000~kg, launches on August with $C3= \mathrm{103~km^2s^{-2}}$. It reaches Halley on September 2060, 4.57~au from the Sun, with a dry mass of 1027~kg. 
An alternative trajectory, with departure on September 2037 and $C3= \mathrm{109~km^2s^{-2}}$, also reaches the comet on September 2060. The rendezvous occurs 4.63~au from the Sun, with a final mass of 916~kg.
These launch energies are high, but not extraordinary. For example, $C3=\mathrm{157~km^2s^{-2}}$ for New Horizons \citep{Stough:2021}.
Both trajectories are characterized by near-ballistic Jupiter-Saturn transfers, followed by a Saturn GA that places the spacecraft almost exactly on the plane required to rendezvous with the comet.
Designing the spacecraft around the mass constraints of the 2037 solution enables two launch attempts. This provides a contingency plan in case the 2036 opportunity is missed.

\section*{Acknowledgments}
R. Flores and E. Fantino received support from the Polar Research Center (PRC) of Khalifa University of Science and Technology (KU) and grant ELLIPSE/8434000533 from Abu Dhabi's Technology Innovation Institute (TII). 
A. Beolchi, C. Pozzi and C. Barbieri acknowledge KU's internal grant CIRA-2021-65/8474000413.

\bibliography{Biblio_Halley}

\end{document}